\documentclass[a4paper,10pt]{article}
\usepackage{amssymb}
\usepackage{amstext}

\title{%
CFT Interpretation\\ of Merging Multiple SLE Traces
}
\author{%
  {\normalsize\sc Annekathrin M\"uller-Lohmann\thanks{{\tt 
    anne@itp.uni-hannover.de}}}\\[0.5cm]
  {\normalsize\slshape Institut f\"ur theoretische Physik}\\[-0.1cm]
  {\normalsize\slshape University of Hannover}\\[-0.1cm]
  {\normalsize\slshape Appelstra\ss e 2}\\[-0.1cm]
  {\normalsize\slshape D-30167 Hannover, Germany}
}
\date{{\small\today}}
\long\def\symbolfootnote[#1]#2{\begingroup%
\def\thefootnote{\fnsymbol{footnote}}\footnote[#1]{#2}\endgroup}

\begin{document}

\maketitle
\enlargethispage{2cm}
\thispagestyle{empty}
\begin{abstract}
{\small 
In this paper we give a physical interpretation of the probability of a Stochastic {\sc L\"owner} Evolution ({\sc SLE}) trace approaching a marked point in the upper half plane, e.\,g.\ on another trace. Our approach is based on the concept of fusion of boundary with bulk or other boundary fields in Conformal Field Theory ({\sc CFT}). In this context we also state a proposal how the results can be extended to situations that happen not sufficiently far away from the boundary. 
}\end{abstract}

\tableofcontents
\section{Introduction}
\subsection{Motivation}
Stochastic {\sc L\"owner} Evolutions ({\sc SLE}) as introduced by O.~Schramm \cite{OdedSchramm} are one parameter families of conformally invariant measures on curves in the plane. For an elaborate introduction, see e.\,g.~\cite{Lawler}. They are described by continuous sets of mappings that satisfy the {\sc L\"owner} differential equation with stochastic driving parameter. This equation has been derived by L\"owner in 1923 in a more general context for non-stochastic 1D real continuous driving functions, allowing for the description of a wider class of 2D curves \cite{Loewner}. 

Taking the driving parameter to be {\sc Brownian} motion has been proven by Schramm to be the only choice for the measure to be conformally invariant, exhibiting reflection symmetry and implying non-self-intersecting curves. This way, the curves can describe the continuum limit of random cluster interfaces connected to the boundary in lattice models of statistical physics at criticality. Actually, this has been rigorously proven for example for the loop-erased random walk ({\sc LERW}) and the uniform spanning tree ({\sc UST}). 

However, for years there has been another way to describe these physical models: (boundary) Conformal Field Theories ({\sc bCFT})s. Unfortunately, in contrast to some {\sc SLE} cases this description is only based on a conjecture. In spite of this fact, the question how the two theories are related to each other arises naturally in this context. Additionally it is generally hoped for results to benefit from an explicit correspondence since in some cases such as percolation \cite{Smirnov:2001-2} it turned out that proofs in one or the other picture are strikingly easier to do.

The community, especially \cite{Bauer:2002tf,Bauer:2006,Friedrich:2004-2,Cardy:2004xs,Bettelheim:2005}, has already been successful in answering this question to some extent; e.\,g.\ observing that the same differential equations arise both in {\sc SLE} and minimal {\sc bCFT} context, heuristically relating the change of boundary condition implicitly fixed by the {\sc SLE} trace to the appeareance of the {\sc bCFT}-fields $\psi_{(1,2)}$ and $\psi_{(2,1)}$. This nontrivial connection is based on lifting the {\sc SLE} stochastic differential equation ({\sc SDE}) of conformal mappings to that of elements of a formal group. Letting these act on appropriate representations of the {\sc Virasoro} algebra, martingales that correspond to null vectors in its {\sc Verma} module can be found \cite{Bauer:2002tf}. This correspondence requires a relation between the speed of the {\sc Brownian} motion and the central charge of the minimal {\sc bCFT}, matching two so-called dual {\sc SLE}s to a minimal {\sc bCFT} model.

Recently, there have been several attempts to relate more than the two boundary fields to objects in (modified versions of) {\sc SLE} \cite{Rasmussen:2004,Moghimi-Araghi:2005} and multiple {\sc SLE} \cite{Bauer:2005jt,Cardy:2003,Dubedat:2004-2}. This has mostly been achieved via the comparison of differential equations or typical scaling behavior, e.\,g.~of the probabilities of visiting small sections on the real or imaginary axis and points in the upper half plane \cite{Bauer:2004,Beffara:2002,Friedrich:2002}, differential equations arising due to the presence of the stress-energy tensor \cite{Cardy:2005} and scaling exponents in the $Q$-states {\sc Potts} model \cite{Cardy:2006}.
\subsection{Outline}
In this work we want to present another contribution to the {\sc SLE-bCFT} relation. Therefore we will review important concepts in the second section and fix the notations. We will start with a short introduction to single and  multiple {\sc SLE} (sections \ref{sec:singleSLE} and \ref{sec:multSLE}), {\sc bCFT} (section \ref{sec:BCFT}), their relationship (section \ref{sec:Obs}) and the short distance expansion in {\sc CFT} (section \ref{sec:fusion}).

The third section is dedicated to {\sc bCFT} and {\sc SLE} probabilities that we want to focus on in this paper. In {\sc bCFT} this addresses the probability of a certain arch configuration occuring in a theory while in {\sc SLE} we are interested in the probability of an {\sc SLE} trace intersecting a small disc of a certain radius.

Afterwards, we will arrive at the key point of this paper in section \ref{sec:scaling}: the behavior of the differential operators of the differential equations that were part of the first step to establish a connection between the two theories. In this part of our paper, we will argue how additional exponents for the small distance behavior can be found by combining the dependency emerging from the small distance expansion of two primary fields with the respective changes in the differential operators.




In the fifth section, we will make use of these results by relating the {\sc SLE} probability to fusion of boundary with bulk fields in section \ref{sec:SLEpoint} and of boundary fields only in section \ref{sec:SLEtrace}. This will lead us to an {\sc SLE} interpretation of correlation functions of other fields than the $\psi_{(1,2)}$ or $\psi_{(2,1)}$-type. Afterwards, we will address the obstacle that the {\sc SLE} probability in question also exhibits an angular dependency for points near the boundary in section \ref{sec:SLEang}.

The last section will be dedicated to another point of view on multiple {\sc SLE}. Briefly reviewing concepts of probability theory in section \ref{sec:radnik}, we will remind the reader that multiple {\sc SLE} can be viewn as ordinary {\sc SLE} weighted by a suitable martingale. This martingale will then be given a probability interpretation in section \ref{sec:probInt}, extending the concept of fusion to this picture in section \ref{sec:fusmart}

The appendix contains the derivation of the limiting behavior of the differential operators stated in section \ref{sec:scaling}. It is straightforward but quite lengthy, hence we restricted ourselves to a heuristic argument in the text, leaving the details of the computations to the appendix.

\section{Basic Definitions and Notations}
\subsection{Single SLE}\label{sec:singleSLE}
Now let us fix some notation. For chordal {\sc SLE}s, the family of {\sc L\"owner} mappings shall be denoted by $(g_t)_{t\in\mathbb{R}^+_0}$ which are defined through the so-called {\sc L\"owner} equation:
\begin{equation}\label{eq:singleSLE}
 \text{d} g_t = \frac{2 \text{d} t}{g_t - \xi_t}\,,\quad  g_0(z) = z\,,\quad g_t(z) = z + \frac{2t}{z} + \mathcal{O}(z^{-2})\text{ for } z \rightarrow \infty\,.
\end{equation}
Here, $\xi_t =\sqrt{\kappa} B_t$ is {\sc Brownian} motion of speed $\kappa$ with $0 \leq \kappa \leq 8$. Additionally we have chosen the standard time parameterization to fix the $SL(2,\mathbb{R}$ invariance such that the half plane capacity is $\text{\sc Hcap}_{\gamma_t} = 2t$. Obviously, $g_t:\mathbb{H}/K_t \rightarrow \mathbb{H}$ where the $K_t$ are the hulls of the so-called traces $\gamma_t :\,= \lim_{\epsilon \rightarrow 0} g_t^{-1}(\xi_t + \mathrm{i}\epsilon)$ that describe the cluster interfaces.  As physical cluster interfaces these traces are continuous and non self-crossing, exhibiting three phases: for $\kappa \leq 4$ they are a.s.\ simple, for $4<\kappa <8$ they are self-touching and for $\kappa \geq 8$ space filling with {\sc Hausdorff} dimension $d_{\gamma_t} = \min \{1+\frac{\kappa}{8},2\}$. The dimension of the {\sc SLE} hull $K_t$ is given by that of its trace for $\kappa < 4$ and by that of the so-called dual {\sc SLE} with parameter $16/\kappa$ for the other cases.

To illustrate the connection to {\sc CFT} we define $h_t (z) :\,= g_t (z) - \xi_t$, satisfying the stochastic differential equation
\begin{equation}
 \text{d} h_t = \frac{2 \text{d} t}{ h_t} - \text{d} \xi_t\,.
\end{equation}
For any time $t$ we associate an element $\mathfrak{g}_{h_t}$ of the germs of holomorphic functions at infinity, $N_{-}$, of the form $z + \sum_{m \leq -1} h_m z^{m-1}$. This satisfies according to {\sc It\^{o}}s formula:
\begin{equation}\label{eq:virSDE}
 \mathfrak{g}_{h_t}^{-1} \cdot \text{d} \mathfrak{g}_{h_t} = \text{d} t\left(-2 l_{-2} + \frac{\kappa}{2} l_{-1}^2 \right) + \text{d} \xi_t l_{-1}\,,
\end{equation}
with $l_{n} = -z^{n+1}\partial_z$. In {\sc CFT}, the $l_n$ correspond to the generators $L_n$ of the {\sc Virasoro} algebra $\mathfrak{vir}$ \cite{Bauer:2002tf}:
\begin{equation}
 [L_m,L_n] = (m-n)L_{n+m} + \frac{c}{12} m(m^2-1) \delta_{m+n,0}\,,
\end{equation}
and it can be shown that there exists a homomorphism $\mathfrak{g}_h \rightarrow G_h$ such that $G_h$ is an operator acting on appropriate representations of $\mathfrak{vir}$, satisfying an equation analogous to (\ref{eq:virSDE}). Now it is easy to see that $G_{h_t} |\psi_{(1,2)}\rangle$ and $G_{h_t} |\psi_{(2,1)}\rangle$ are local martingales if the $\psi$s are primary fields of weights $h_{(1,2)} = \frac{6-\kappa}{2\kappa}$ or $h_{(2,1)} = \frac{3\kappa-8}{16}$. Indeed, this is equivalent\symbolfootnote[3]{Note that in \cite{Rushkin:2007} it has been shown that this identification is only true modulo a phase that accounts for the branch points of {\sc CFT} correlation functions. However, to keep things simple, we will not concentrate on this subtlety in this paper.} to saying that these fields have a degenerate descendent on level two:
\begin{equation}
 \left( -2 L_{-2} + \frac{\kappa}{2} L_{-1}^2 \right) |\psi_h\rangle = 0\,,
\end{equation}
if the following relation between the central charge of the {\sc CFT} $c$ and the speed of the {\sc Brownian} motion $\kappa$ holds \cite{Bauer:2002tf}:
\begin{equation}\label{eq:c}
 c = \frac{(3 \kappa-8)(6-\kappa)}{2\kappa} \leq 1\,.
\end{equation}
\subsection{Multiple SLE}\label{sec:multSLE}
Multiple {\sc SLE} following \cite{Graham:2005}, based on \cite{Bauer:2005jt,Dubedat:2004-2}, can be regarded as $m$ single chordal {\sc SLE}s in the same domain. For consistency, some requirements such as conformal invariance, reparameterization invariance and absolute continuity have to be made. Allowing local growth at $m$ tips in the upper half plane results in a modified {\sc L\"owner} mapping $G_t$ that describes $m$ single {\sc SLE}s in a single equation:
\begin{equation}\label{eq:multipleLoewner}
\text{d} g^i_t(z) = \frac{2 c_t^i}{g_t^i - w_t^i} \text{d}t \quad \text{for } i=1,\ldots,m \quad \rightarrow \quad\text{d} G_t(z) = \sum_{i=1}^m \frac{2 a_t^i \text{d} t}{G_t(z) - x_t^i}\,.
\end{equation}
Defining $G_t =: H_t^i \circ g_t^i$, we can specify the relationship between the old and the new driving parameters $x_t^i = H_t^i(w_t^i)$ and time parameterisations $a_t^i = H_t^i{}'(w_t^i)^2 c_t^i$. Loosly speaking, $H_t^i$ is the mapping that removes the remaining $n-1$ {\sc SLE} traces from the setting after the action of $g_t^i$.

The conditions lead us to the $\kappa$-relation: all involved speeds $\kappa_i$ of the $m$ {\sc SLE}s should correspond to the same central charge. From (\ref{eq:c}) we can deduce that for all $i,j \in \{1,\ldots,m\}$ we have $\kappa_i = \kappa_j$ or $\kappa_i = 16 / \kappa_j $. 

Another implication is a change of the measure that leads to a change of the drift term from purely {\sc Brownian} motion to:
\begin{equation}\label{eq:driftchange}
\text{d} w_t^i 
= \sqrt{\kappa} \text{d} B_t \rightarrow \text{d} x_t^i 
= \sqrt{\kappa_i}\text{d}  B_t^i + \kappa_i a_t^i \partial_{x_t^i} \log Z [x_t] \text{d} t + \sum_{k \neq i} \frac{2 a_t^i}{x_t^i - x_t^k} \text{d}t\,,
\end{equation}
where $Z[x_t]$ is usually being interpreted as the partition function of the corresponding {\sc bCFT}.

Enforcing conformal invariance and absolute continuity yields an exponential change of the measure given by the {\sc Radon}-{\sc Nikodyn} derivative according to {\sc Girsanov}'s theorem. Hence, similarly to the interpretation of {\sc SLE}($\kappa,\vec{\rho}$) \cite{Werner:2004}, we can interpret this {\sc SLE} again as ``usual'' {\sc SLE} only weighted by another local bounded martingale \cite{Graham:2005}. Here, the martingale is given by
\begin{eqnarray} \label{eq:martingale}
M_t&:\,=&\frac{Z[x_t]}{Z[x_0]} \prod_{i=1}^m\left(H_t^i{}'(w_t^i)\right)^{h_i} \exp \left( \frac{c_i}{6} \int_0^t \mathcal{S} H_s^i(w_s^i) c_s^i \text{d} s \right) \nonumber\\
&\quad& \quad \quad\quad\quad \exp \left( -\int_0^t \frac{1}{Z[x_s]} \mathcal{D}_{-2}^m(x_s^i) Z[x_s] a_s^i \text{d} s\right)\,,
\end{eqnarray}
where $\mathcal{S}$ denotes the {\sc Schwarz}ian derivative and $\mathcal{D}_{-2}^m(x_s^i) $ is given by
\begin{equation}
 \mathcal{D}_{-2}^m(x_s^i) = \frac{\kappa_i}{2} \partial^2_{x_s^i}
-2\sum_{k\neq i}\left( \frac{h_k}{(x_s^k -x_s^i)^2} - \frac{1}{(x_s^k -x_s^i)} \partial_{x_s^i}\right)\,.
\end{equation}
\subsection{Boundary CFT Revisited}\label{sec:BCFT}
The goal of this side trip is to point out the subtleties of {\sc bCFT} on the upper half plane since it is usually dealt with as a chiral theory on the full complex plane which forces us to think about how to extend the {\sc L\"owner} equation from $\mathbb{H}$ to $\mathbb{C}$, too. Additionally, the boundary behavior of bulk fields gets affected by the formalism.

In {\sc bCFT}, due to the boundary conditions that have to be imposed one way or the other, the behavior of the primary fields under conformal transformation changes. Hence, the way we have to treat observables that are given by products of primary fields $\phi(z,\bar{z})$ with holomorphic and antiholomorphic coordinate dependencies differs from that of the boundary fields $\psi(x)$; 
\begin{eqnarray}
 \phi(z,\bar{z}) &\rightarrow& \phi\left(f(z),f(\bar{z})\right) =  (f'(z))^{-h} (f'(\bar{z}))^{-\bar{h}} \ {}^f \phi(z,\bar{z})\,,\\
\psi(x) &\rightarrow&  \psi\left(f(x)\right) =  (f'(x))^{-h}\ {}^f \psi(x)\,.
\end{eqnarray}
wherein $f$ denotes the functional change of the field due to the conformal transformation.

In conformal field theory on the full complex plane, we can regard the holomorphic and antiholomorphic coordinates, $z$ and $\bar{z}$ as independent variables since any conformal transformation factors into the two parts due to the {\sc Cauchy}-{\sc Riemann} differential equations. Any variation of a correlation function with respect to a conformal transformation given by $z \mapsto z + \epsilon (z)$ and $\bar{z} \mapsto \bar{\epsilon}(\bar{z})$ can be written as
\begin{equation}
 \delta_{\epsilon,\bar{\epsilon}} \langle X \rangle_{\mathbb{C}} = - \frac{1}{2 \pi \text{i}} \oint_{\mathcal{C}} \text{d}z \, \epsilon(z) \langle T(z) X \rangle + \frac{1}{2 \pi \text{i}} \oint_{\mathcal{C}} \text{d}\bar{z} \, \bar{\epsilon}(\bar{z}) \langle \bar{T}(\bar{z}) X \rangle\,.
\end{equation}
with the counterclockwise contour $\mathcal{C}$ including all the (anti-)holomorphic positions of the primary fields in $X$.

However, in boundary {\sc CFT} on the upper half plane $\overline{\mathbb{H}}$ including the real axis, we are confined to those conformal transformations that leave the real axis, i.\,e.\ the boundary, invariant. This gives us the constraint $\epsilon (x) = \bar{\epsilon}(x)$ and $T(x) = \bar{T}(x)$ for any $x\in\mathbb{R}$. Taking a look at a correlation function trying to seperate the holomorphic and antiholomorphic parts again, we see that we have a non-vanishing contribution from the boundary:
\begin{eqnarray}
\delta_{\epsilon} \langle X \rangle_{\overline{\mathbb{H}}} &=& - \frac{1}{2 \pi \text{i}} \oint_{\mathcal{C}^+} \text{d}z \, \epsilon(z) \langle T(z) X \rangle -  \frac{1}{2 \pi \text{i}} \int_{-\infty}^\infty \text{d}x \, \epsilon(x) \langle T(x) X \rangle \,, \label{eq:varUHP}\\
 \delta_{\bar{\epsilon}} \langle X \rangle_{\overline{\mathbb{H}}} &=& \frac{1}{2 \pi \text{i}} \oint_{\mathcal{C}^+} \text{d}\bar{z} \, \bar{\epsilon}(\bar{z}) \langle \bar{T}(\bar{z}) X \rangle + \frac{1}{2 \pi \text{i}} \int_{-\infty}^\infty \text{d}x \, \bar{\epsilon}(x) \langle \bar{T}(x) X \rangle\,.\label{eq:varLHP}
\end{eqnarray}
where $\mathcal{C}^+$ indicates a counterclocise contour in the upper half plane $\overline{\mathbb{H}}$ (including the real axis) encircling all (anti-)holomorphic positions of the primary fields in $X$.

Obviously, due to the boundary conditions, both boundary terms give exactly the same contribution -- the two terms are no longer independent. From now on we will set $\bar{z} = z^*$ and consider the antiholomorphic quantities, e.\,g.\ $\bar{T}(\bar{z})$, as being the analytic continuation of the holomorphic quantities, e.\,g.\ $T(z)$ in the lower half plane \symbolfootnote[3]{This is possible since $\bar{T}(z^*)$ for $\mathfrak{Im}(z) <0$ is holomorphic and $T(z) = \bar{T}(\bar{z})$ on the real axis.}. Thus we arrive at only one set of {\sc Virasoro} generators by taking advantage of $\delta_{\bar{\epsilon}} \langle X \rangle_{\overline{\mathbb{H}}} = 0$ and reexpressing the boundary integral in one of the equations (\ref{eq:varUHP}) and (\ref{eq:varLHP}). In this picture, the bulk fields depending on holomorphic and antiholomorphic coordinates become two seperate fields, one being the ``mirror image'' of the other $\phi (w,\bar{w})_{\overline{\mathbb{H}}}= \phi (w)_{\overline{\mathbb{H}}}\otimes\phi(\bar{w})_{\mathbb{C} / \mathbb{H}} = \phi (w)_{\overline{\mathbb{H}}}\phi(w^*)_{\mathbb{C} / \mathbb{H}}$. Effectively, we are using $2n$ holomorphic degrees of freedom in this picture instead of $n$ holomorphic and $n$ antiholomorphic ones with a boundary condition:
\begin{eqnarray}
\delta_{\epsilon} \langle X \rangle_{\mathbb{C}_b} &=& - \frac{1}{2 \pi \text{i}} \oint_{\mathcal{C}^+} \text{d}z \, \epsilon(z) \langle T(z) X \rangle + \frac{1}{2 \pi \text{i}} \oint_{\mathcal{C}^-} \text{d}z \, \epsilon(z) \langle \bar{T}(z) X \rangle \,.\label{eq:varCP}
\end{eqnarray}
This point of view suggests that by only considering conformal transformations $f(z)$, $f: D\subset \overline{\mathbb{H}}\rightarrow\overline{\mathbb{H}}$ that preserve the boundary, we have to modify the behaviour of the primary bulk and boundary fields under these transformations:
\begin{eqnarray}
 \phi(z,\bar{z}) &\rightarrow& \phi\left(f(z),f(\bar{z})\right) = (f'(z))^{-h} (f^*{}'(z)^*)^{-\bar{h}}\ {}^f \phi(z,\bar{z})\,, \label{eq:bulkbeh}\\
\psi(x) &\rightarrow& \psi\left(f(x)\right) =  (f'(x))^{-h}\ {}^f \psi(x)\,.\label{eq:boundarybeh}
\end{eqnarray}
Equation (\ref{eq:bulkbeh}) is precisely the point which is usually ignored in the literature, as well as the doubling of points when going from the usual picture with chiral and antichiral halves on the upper half plane to the only-chiral theory on the full complex plane. 

Hence, any differential equation resulting from degenerate fields involves holomorphic as well as antiholomorphic coordinates, e.\,g.\ the level two differential equation arising in minimal {\sc CFT}s that imposes constraints on correlation functions including $\psi_{(2,1)}$ or $\psi_{(1,2)}$ boundary field. 

If we choose the {\sc L\"owner} mapping $f(z) = g_t(z)$, we will have to take terms including the antiholomorphic coordinates into account if we want to include a truely local observable or want to study boundary behavior which is usually done by taking a look at the limit
\begin{equation}
 \lim_{z\rightarrow \bar{z}} \phi(z)\phi(\bar{z})\,,
\end{equation}
and inserting the {\sc OPE} (see section \ref{sec:fusion}) for the product of the chiral and antichiral half of the local field $\phi(z,\bar{z})$. In that case, we have to extend the {\sc L\"owner} mapping to the lower half plane which should be done by mirroring as well. Hence we define for $ z,\bar{z} \in \overline{\mathbb{H}}$ or, equivalently $w \in \mathbb{C}$:
\begin{eqnarray}
z \mapsto g_t(z) \quad &\text{and}& \quad
\bar{z} \mapsto g_t(\bar{z}) \quad \text{with b.c. or}\\
w \mapsto g_t(w) &:\,=& \left\lbrace
\begin{array}{ll}
g_t(w) & \text{ for } \mathfrak{Im}(w) > 0\,,\\
g_t^*(w^*) & \text{ for } \mathfrak{Im}(w) < 0\,.
\end{array}
\right.
\end{eqnarray}
\subsection{SLE Martingales and Physical Quantities in BCFT}\label{sec:Obs}
Following the nice introductions \cite{Bauer:2005jt,Santachiara:2007,Bettelheim:2005,Rushkin:2007} to the connection between {\sc SLE} and minimal models of {\sc bCFT}, we will give a short overview on how martingales in multiple {\sc SLE} are connected to correlation functions in {\sc bCFT}. 

On the one hand, it is quite obvious from the definition of martingales in the theory of probability that a quantity $M_t$ which satisfies all conditions for martingales can be referred to as ``conserved'', i.\,e.\ not changing its expectation value in time, or, more mathematically: $\mathbb{E} (M_t) = \mathbb{E}(M_0)$ for all times $t$ which means nothing else than $\frac{\text{d}}{\text{d}t} \mathbb{E}(M_t) = 0$. Hence, if we assume that observables $O^{\text{\sc stat.P.}}_t$ should be conserved quantities, we can also say that $\mathbb{E}(M_t) = \mathbb{E}(O^{\text{\sc SLE}}_t)$. 

On the other hand, in {\sc CFT} it is common believe that all physically interesting quantities are given in terms of correlation functions of the fields in the theory -- i.\,e.\ the possible amplitudes of the scattering matrix. Within the theories of minimal models, these correlation functions $\mathcal{O}^{\text{\sc CFT}}$ are known to obey certain differential equations $\mathcal{D} \mathcal{O}^{\text{\sc CFT}} = 0$, e.\,g.\ due to the existence of singular states or {\sc Ward} identities. 

Hence, the key point is basically to realize that there are correlation functions in {\sc bCFT} and martingales in {\sc SLE} that describe the same observables in statistical physics. Once we accept this, we can show that they are actually described by the same differential equation, i.\,e.\ $\frac{\text{d}}{\text{d}t} \mathbb{E} (O^{\text{\sc SLE}}_t)$  with $O_t^{\text{\sc SLE}} \rightarrow O_t^{\text{\sc CFT}}$ will turn out to be the null vector equation of {\sc bCFT}: $\mathcal{D} \mathcal{O}^{\text{\sc CFT}} = 0$. 

In the following we will illustrate how the probabilistic description of {\sc SLE} martingales is connected to the statistical physics language. Unfortunately, we can not be rigorous here since the connection between the respective statistical physics models and {\sc bCFT} is still only based on a conjecture. Hence the point where we will take the continuum limit is nly based on an assumption which is quite widely believed but has never been proven.

Nevertheless let us have a look at a statistical mechanics model with boundary, e.\,g.\ defined on $\mathbb{H}$, whose continuum limit at the critical point can be described by a {\sc bCFT}\symbolfootnote[4]{As said above, we are aware of the fact that his part still lacks a proper proof but we will nevertheless assume its correctness here.}. Hence, at least for most observables $O^{\text{\sc stat.P.}}$ we assume that their expectation values will be described by {\sc bCFT} correlation functions. We will allow for a finite but large set of possible states $S$ to whose elements normalized {\sc Boltzmann} weights $w(s)$ with probability $\textbf{P} = w(s) / Z$ are attatched where $Z=\sum_{s\in S}w(s)$ denotes the partition function. Hence we can state the expectation value of observables described by random variables $O^{\text{\sc stat.P.}}\,:\,S\rightarrow \mathbb{H}$ as:
\begin{equation}
 \mathbb{E}(O^{\text{\sc stat.P.}}) = \mathcal{O}^{\text{\sc stat.P.}} = \frac{1}{Z} \sum_{s\in S} O^{\text{\sc stat.P.}}(s) w(s)\,.
\end{equation}
Now we want to restrict ourselves to a special situation, leading to a conditioned expectation value, i.\,e. an expectation value with respect to the knowledge of the situation up to some time $t$. This will allow us to relate the statistical physics expectation value to the one of {\sc SLE} by taking subsets of $S$ that correspond to interfaces $\gamma_t$ between the boundary and an interior point. Therefore we define $(S_\alpha)_{\alpha \in I}$ to be a collection of disjoint subsets of $S$ whose union is again $S$ to get a sigma algebra $\mathcal{F} = \lbrace \bigcup_{\alpha \in I'} S_\alpha \, : \, I' \subset I\rbrace$. This induces a filtration, i.\,e.\ an increasing family $(\mathcal{F}_t)_{t \geq 0}$ of sigma algebras, $\lbrace \emptyset, S \rbrace \subset \mathcal{F}_s \subset \mathcal{F}_t $ for all $0 \leq s < t$. Hence the partition function for this special situation is given by
\begin{equation}
 Z_\alpha^{(t)} = \sum_{s \in S_\alpha^{(t)}} w(s)\,.
\end{equation}
and the conditional expectation value is:
\begin{equation}
 \mathcal{O}^{\text{\sc SLE}}_t = \mathbb{E}(O^{\text{\sc SLE}} | \mathcal{F}_t ) = \sum_{\alpha \in I_t} \left(\frac{1}{Z_\alpha^{(t)}} \sum_{s\in S_\alpha^{(t)}} O^{\text{\sc stat.P.}}(s) w(s) \right) \textbf{1}_{S_\alpha^{(t)}}\,.
\end{equation}
Obviously, by definition, this is a martingale. In the following we will use the shorthand notation $O^{\text{\sc SLE}}_t = O^{\text{\sc SLE}} | \mathcal{F}_t $.

In our case, the special situation shall be that of $m$ interfaces $\gamma_t^i$ emerging at the points $x_0^i$ on the boundary and growing (not intersecting each other) into the upper half plane. Our filtration therefore is given by $\mathcal{F}_t = \sigma( \gamma_{t'}^i \,:\, 0 \leq t' \leq t, \, i=1,\ldots,m )$. As always, we define $G_t: \mathbb{H}/ \bigcup_{i=1}^m K_t^i \rightarrow \mathbb{H}$.

Hence, in the continuum limit we follow the general assumption that we can identify the statistical mechanics observable that we have already shown to be an {\sc SLE} martingale with a {\sc bCFT} correlation function:
\begin{equation}
 \mathcal{O}_t^{\text{\sc SLE}} \rightarrow \frac{\mathcal{O}_{\mathbb{H}_t}^{\text{\sc CFT}} }{\left\langle \textbf{1} \right\rangle_{\mathbb{H}_t}^{\text{\sc CFT}}} = \frac{\left\langle \psi(\infty), ^{G_t}O (\lbrace (z_t^k,\bar{z}_t^k)\rbrace)\prod_{l=1}^m \psi (x_t^l) \right\rangle_{\mathbb{H}}^{\text{\sc CFT}}}{\left\langle \psi(\infty), \prod_{l=1}^m \psi (x_t^l) \right\rangle_{\mathbb{H}}^{\text{\sc CFT}}}\,,
\end{equation}
for some observable $O (\lbrace (z_t^k,\bar{z}_t^k)\rbrace) = \prod_{k=1}^n \phi(z_t^k,\bar{z}_t^k)$ with the upper index $G_t$ denoting the action of the {\sc L\"owner} mapping on the observable. Note that we did not write out the {\sc Jacobi}an factors for the also transformed boundary fields -- they cancel in the numerator and denominator.

Now we will explicitly compute the {\sc It\^{o}} derivative to arrive at the relationship with null vector conditions of {\sc bCFT}. Therefore we will state some intermediate results.

In minimal {\sc bCFT}, the fields $\phi(z,\bar{z})$ are primary fields and thus behave under a conformal map $G_t$ such as the multiple {\sc L\"owner} mapping as follows:
\begin{equation}
^{G_t} \phi_j (z_j,\bar{z}_j) = \left(G_t'(z_j) \right)^{h_j}\left(G_t'(z_j)^* \right)^{\bar{h}_j} \phi_j \left(G_t(z_j),G_t(z_j)^*\right)\,,
\end{equation}
where $h_j$ is the weight of the field $\phi_j$ and $z_j = z_0^j$ the coordinate in the initial domain $\mathbb{H}_0 = \mathbb{H}$. Thus we can use this to compute the variation of the bulk fields $\text{d} \left( ^{G_t} \phi_j(z_j,\bar{z}_j)\right) = ^{G_{t+\text{d}t}} \phi_j (z_j,\bar{z}_j) - ^{G_t} \phi_j (z_j,\bar{z}_j)$ introducing the simplified notation $G_t(z_j) = z_t^j$:
\begin{eqnarray}
\text{d}\left( ^{G_t} \phi_j(z_j,\bar{z}_j)\right) &=& \text{d} \left(\left(G_t' \right)^{h_j} \left(G^*_t{}' \right)^{\bar{h}_j} \phi_j \left(z_t^j,z_t^*{}^j\right)\right)\nonumber\\
&=& h_j \left( G_t' \right)^{h_j-1} \text{d} G_t'  \left(G^*_t{}' \right)^{\bar{h}_j} \phi_j \left(z_t^j,z_t^*{}^j\right) \nonumber\\
&\quad& + \,\bar{h}_j \left( G_t^*{}' \right)^{\bar{h}_j-1} \text{d} G_t^*{}'  \left(G_t' \right)^{h_j} \phi_j \left(z_t^j,z_t^*{}^j\right) \nonumber\\
&\quad& + \left( G_t'\right)^{h_j} \left(G^*_t{}' \right)^{\bar{h}_j}\text{d}\phi_j \left(z_t^j,z_t^*{}^j\right)
\end{eqnarray}
Switching the order of the derivatives with respect to $z_j$ and $t$, we can use the {\sc L\"owner} differential equation to compute $\text{d} G_t'$ and $\text{d} G_t^*{}'$. Additionally, we use the chain rule for $\text{d}\phi_j \left(z_t^j, z_t^*{}^j\right)$:
\begin{eqnarray}\label{eq:dphi}
\text{d}\left( ^{G_t} \phi_j(z_j,\bar{z}_j)\right) &=& -\sum_{i=1}^m 2 \left\lbrace\left( \frac{h_j}{\left(z_t^j - x_t^i\right)^2} - \frac{1}{ \left( z_t^j - x_t^i\right)} \partial_{z_t^j} \right)\right. \\
&\quad& 
\left. -\left( \frac{\bar{h}_j}{\left(z_t^*{}^j - x_t^i\right)^2} - \frac{1}{\left( z_t^*{}^j - x_t^i\right)} \partial_{z_t^*{}^j}\right)
\right\rbrace \text{d} t\, ^{G_t} \phi_j \left(z_j, z_j^*\right)\,.\nonumber
\end{eqnarray}
As the boundary fields depend on stochastic variables, we have to apply {\sc It\^o}'s rule. Note that the off-diagonal contributions vanish due to $\text{d}t^2 = \text{d}B_t^i \text{d}t = 0$ and $\text{d}B_t^i \text{d}B_t^j = \delta_{ij} \text{d}t$:
\begin{equation}\label{eq:dpsi}
\text{d} \psi_i (x_t^i) = \partial_{x_t^i}\psi (x_t^i) \text{d} x_t^i + \frac{\kappa}{2} \partial^2_{x_t^i} \psi(x_t^i) \text{d} t\,.
\end{equation}
With the help of equations (\ref{eq:dphi}) and (\ref{eq:dpsi}), it can be shown explicitly that
\begin{equation}
 \mathcal{O}_t^{\text{\sc CFT}} = \frac{\left\langle \psi(\infty), O(\lbrace (G_t(z_j),G_t^*(z_j^*)\rbrace) \prod_{l=1}^m \psi (x_t^l) \right\rangle}{\left\langle \psi(\infty),\prod_{l=1}^m \psi (x_t^l)\right\rangle} \prod_{k=1}^n G_t '(z_k)^{h_k} G_t^*{}'(z_k^*)^{\bar{h}_k}
\end{equation}
is (not only by definition) a closed martingale, obeying $m$ differential equations for $i=1,\ldots,m$. This can be shown via taking the time derivative of $\mathcal{O}_t^{\text{\sc SLE}}$, using the {\sc L\"owner} mapping and the quotient rule of the {\sc It\^{o}} calculus to get:
\begin{eqnarray}\label{eq:expectation}
\frac{\text{d}}{\text{d}t} \mathbb{E} \left(O_t^{\text{\sc SLE}} \right) &=&
\left[ \frac{\kappa}{2} \partial^2_{x_t^i} 
-2\sum_{l\neq i=1}^m \left( \frac{h_l}{\left(x_t^l - x_t^i\right)^2} - \frac{1}{\left( x_t^l - x_t^i\right)} \partial_{x_t^l} \right)\right.\nonumber\\
&\quad& -2 \sum_{j=1}^n \left\lbrace\left( \frac{h_j}{\left(z_t^j - x_t^i\right)^2} - \frac{1}{\left( z_t^j - x_t^i\right)} \partial_{z_t^j} \right)\right.\nonumber\\
&\quad&\quad\quad\quad \left.\left. -\left( \frac{\bar{h}_j}{\left(z_t^*{}^j- x_t^i\right)^2} - \frac{1}{\left( z_t^*{}^j - x_t^i\right)} \partial_{z_t^*{}^j}
\right)
\right\rbrace
\right]
\mathcal{O}_t^{\text{\sc CFT}}\nonumber\\
&=\,:& \mathcal{D}_{-2}^{m,n} \mathcal{O}_t^{\text{\sc CFT}}(x_t^i,\{x_t^l\}_{l\neq i},\{z_t^k,z_t^*{}^k\})\nonumber\\
&=&0\,.
\end{eqnarray}
Now, following \cite{Bauer:2005jt}, we will introduce $m$ functions $x^l$ and $n$ functions $z^k,z^*{}^k$, given by $t \mapsto x_t^l$ and $t \mapsto z_t^k$, $t \mapsto z_t^*{}^k$, respectively. In this notation, equation (\ref{eq:expectation}) becomes
\begin{equation}
 \frac{\text{d}}{\text{d}t} \mathbb{E} \left(O_t^{\text{\sc SLE}} \right) = \mathcal{D}_{-2}^{m,n}\mathcal{O}^{\text{\sc CFT}} (x^i,\{x^l\}_{l\neq i},\{z^k,z^*{}^k\})
\end{equation}
which shall be interpreted as a functional expression: the {\sc r.h.s.} is true for any coordinate that the {\sc L\"owner} images of the traces $\gamma$ can assume, hence it does not depend on the specific point in time we choose on the {\sc l.h.s}. This is not suprising at all, since $\mathcal{O}_t^{\text{\sc SLE}}$ is an {\sc SLE} martingale, hence we expect the time derivative of its expectation value to vanish anyway. 
In words this relation shows that the martingale condition in {\sc SLE} for statistical physics observables translates to null vector differential equations obeyed by the same observables expressed via the {\sc CFT} correlation functions. 
Heuristically, this shows that, indeed, in the continuum limit, both models describe the same physics and hence some quantities may be related to each other.

The inclined reader may also find another point of view on the relationship between the {\sc SLE} and {\sc CFT} form of the differential equation in \cite{Rushkin:2007}.



Note that if we chose $m=1$, we would get the single {\sc SLE} case back since there are only two boundary fields in that case -- $\psi_{(2,1)}(0)$ and $\psi_{(2,1)}(\infty)$. Hence the partition function would reduce to the free case since (naively)
\begin{equation}
\left\langle \psi_{(2,1)}(\infty),\psi_{(2,1)}(\xi_t) \right\rangle \propto \lim_{x_\infty \rightarrow \infty} (\xi_t-x_\infty)^{-2h_{(2,1)}}
\end{equation}
is usually set to a constant. Hence we would have no differential equations for the correlation function of the boundary fields alone.

\subsection{Fusion and the OPE in CFT}\label{sec:fusion}
In order to be able to study and describe the behavior of quantities approaching each other properly, we have to use the short distance expansion of products of primary fields in {\sc CFT}.

In minimal {\sc CFT}, the primary fields correspond to heighest weights in the {\sc Ka\u{c}} table, characterized by their {\sc Ka\u{c}} labels $(r,s)$. The so-called fusion rules tell us which primaries and descendants are involved in the short distance product (``operator product expansion'' ({\sc OPE})) of two given fields\cite{DiFrancesco:1997nk}:
\begin{equation}
 \phi_0 (z) \phi_1(w) =
\sum_h g_{(r',s')} 
\sum_Y (z-w)^{h - h_0 -h_1 +|Y|} \beta_{Y,(r',s')} L_{-Y} \phi_h (w)\,,
\end{equation}
where $Y = \{ k_1,k_2,\ldots k_n\}$, $k_1\geq \ldots \geq k_n$, $|Y| = \sum_{i=1}^n k_i$ and $ g_{(r',s')} $ the coefficient of the three-point function of the involved fields of weights $h_0 = h_{(r_0,s_0)}$, $h_1 = h_{(r_1,s_1)}$ and $h = h_{(r',s')}$.

In principle, if the two fundamental fields $\phi_{(2,1)}$ and $\phi_{(1,2)}$ are present in a theory, all other fields may be generated by consecutive fusion of a suitable number of copies. For our purposes, it suffices to know that
\begin{eqnarray}
 \phi_{(1,2)} \times \phi_{(1,2)} =  \phi_{(1,1)} +  \phi_{(1,3)}\,,\\
 \phi_{(2,1)} \times \phi_{(2,1)} =  \phi_{(1,1)} +  \phi_{(3,1)}\,.
\end{eqnarray}
Thus if the {\sc SLE} trace tips correspond to boundary fields of type $\psi_{(2,1)}$ or $\psi_{(1,2)}$, it should in principle be possible to find quantities that correspond to the other entries of the {\sc Ka\u{c}} table by investigating merging traces in multiple {\sc SLE}. Hence, we will focus on the behavior of {\sc SLE} and {\sc CFT} quantities in which two coordinates approach each other.
%
\section{Probabilities in BCFT and SLE}
In the first part of this section we will give an heuristic approach to probabilities related to the partition function in {\sc bCFT}. 

In the second part we will give a short review on an important probability of a special situation in {\sc SLE}: the answer to the question asking how probable it is for the {\sc SLE} trace to hit a small ball in the upper half plane, hence being in the vicinity of a marked point. Of course, this probability has been derived for single {\sc SLE}s. However, multiple {\sc SLE} can always be viewn as $m$ single {\sc SLE}s if considering small time steps or large distances between the curves. Thus we think that it is justified to use the formula originally derived only for single {\sc SLE}s for our purposes, too.
\subsection{Partition Functions and Probabilities in BCFT}\label{sec:probBCFT}
In boundary {\sc CFT}, the partition function of a system with boundary conditions $\alpha_1, \ldots, \alpha_m$ changing at the positions $x_1,\ldots,x_m \in \mathbb{R}$ is given by \cite{DiFrancesco:1997nk}
\begin{equation}
Z_{\alpha_1, \ldots, \alpha_m} = \left\langle\psi(\infty), \prod_l^m \psi_{\alpha_l,\alpha_{l+1}}(x_l)  \right\rangle Z_{\text{free}}\,.
\end{equation}
Interpreting the partition function of the free system, $Z_{\text{free}}$, as a measure of the total number of states and the partition function of the system with fixed boundary conditions, $Z_{\alpha_1, \ldots, \alpha_m}$, as a measure for the interfaces due to the changes in the boundary conditions, we see that the fraction
\begin{equation}
\frac{Z_{\alpha_1, \ldots, \alpha_m}}{Z_{\text{free}}} =: \textbf{P}_{\gamma_1,\ldots,\gamma_m}
\end{equation}
is the fraction of interfaces $\gamma_1,\ldots,\gamma_m$ fulfilling the given boundary conditions among all possible interfaces of the full system \cite{Friedrich:2004}. Therefore switching to the special situation introduced in section \ref{sec:Obs}, we will consider $\textbf{P}_{\gamma_t^1,\ldots,\gamma_t^m}$ as the probability that $m$ given paths emerge from the boundary points $x_t^1,\ldots, x_t^m$, i.\,e.:
\begin{equation}
 \textbf{P}_{\gamma_t^1,\ldots,\gamma_t^m} = \left\langle \psi(\infty) ,\prod_l^m \psi_{\alpha_l,\alpha_{l+1}}(x_t^l) \right\rangle\,,
\end{equation}
which is the same probability now conditioned on already having $m$ interfaces $\gamma_{(0,t]}$ in $\mathbb{H}$, growing further to infinity.

In \cite{Bauer:2005jt} this probability has been shown to be a martingale with respect to the multiple {\sc SLE} measure obeying $m$ differential equations. Thus for $i=1,\ldots,m$ we have:
\begin{equation}
 \left( \frac{\kappa_i}{2} \partial_{x_t^i}^2 - 2 \sum_{l\neq i=1}^m \left( \frac{h_l}{(x_t^l - x_t^i)^2} - \frac{1}{x_t^l - x_t^i} \partial_{x_t^l} \right) \right) \textbf{P}_{\gamma_t^1,\ldots,\gamma_t^m} = 0\,,
\end{equation}
with $h_i,h_l\in\{h_{(1,2)},h_{(2,1)}\}$ for the setting to be described by {\sc SLE}, too.

From this follows, that
\begin{equation}
 \frac{Z[x_t]}{Z[x_0]} = \frac{\textbf{P}_{\gamma_t^1,\ldots,\gamma_t^m}}{\textbf{P}_{\gamma_0^1,\ldots,\gamma_0^m}}
 %
%
\,,
\end{equation}
where $\textbf{P}_{\gamma_0^1,\ldots,\gamma_0^m}$ is just the probability to chose $m$ specific starting points for our {\sc SLE}. Hence the ratio of the partition function of time $t$ to time $0$ gives us the probability that we have $m$ {\sc SLE} traces up to time $t$ that started somewhere on the boundary, i.\,e.\ the arch configuration where we have $m$ archs not having paired up to time $t$.



\subsection{The SLE Probability of Intersecting a Disc}\label{sec:SLEprob}
In \cite{Beffara:2002}, the probability of an {\sc SLE}$_\kappa$ trace $\gamma$, ${0<\kappa<8}$, intersecting a disc $\mathcal{B}_\epsilon (z_0)$ of radius $\epsilon$ centered at a point $z_0$ in the upper half plane is derived based on an idea of Oded Schramm. It extends the result of chapter 7.4 of \cite{Lawler}:
\begin{equation}\label{eq:Pepsz0}
\textbf{P}_{\epsilon,z_0} := \textbf{P}\left(\gamma_{(0,\infty]} \cap \mathcal{B}_\epsilon (z_0) \neq \emptyset \right)
\asymp \left( \frac{\epsilon}{\mathfrak{Im}(z_0)} \right)^{2 - d_{\gamma}} \left( \sin \alpha(z_0) \right)^{8/\kappa-1}\,,
\end{equation}
where $d_{\gamma} = \text{max}\{2,1+\kappa/8\}$ is the dimension of the {\sc SLE} path and $\alpha(z_0)$ the angle between the real axis and the vector pointing to $z_0$. 

However, in the following we want to study the merging of {\sc SLE} traces. Since they are conditioned not to intersect with each other, they can only come close to each other on their ``outer parts'', i.\,e.\ the hulls $K_t$. Thus for $4 < \kappa < 8 $ we will have to replace $d_{\gamma}$ by $d_{K}$, the dimension of the {\sc SLE} hull $K_t$, which is given by $ 1 + \frac{2}{\kappa}$ ($d_{\gamma}$ for an {\sc SLE} with speed $16/\kappa$). 

Note that the angular dependency on $\alpha(z_0)$ becomes dominant for points near the boundary  but can be neglected for those sufficiently far away. Therefore we will split the discussion of the dependency on the distance $\epsilon$ and the angle $\alpha(z_0)$.
\section{The Effect of Fusion on the Scaling Properties}\label{sec:scaling}
This section is dedicated to the behavior of the differential operators acting on the correlation functions in {\sc bCFT} or martingales in {\sc SLE}. A short heuristic introduction to the quite exhausting computational part can be found in the appendix \ref{app:fusion}. 

From here on we will make use of the interpretation introduced at the end of section \ref{sec:Obs}: we will replace the dependency on the coordinates $x^l_t,z^k_t$ by the functional dependency on $x^l, z^k$, respectively.
\subsection{The L\"owner Equation Differential Operators}\label{sec:scaling1}
For any $i=1,2,\ldots,m$ and $h_j,\bar{h}_j \neq 0$ (if $h_j = 0$ or $\bar{h}_j = 0$, the corresponding term in the sum just vanishes), let us define 
\begin{eqnarray}\label{eq:bulkboundaryop}
\mathcal{D}^{m,n}_{-2}(x^i,\{x^l\},\{(z^k,\bar{z}^k)\}) \!\!\!\!&:\,=&\!\!\!\! \frac{\kappa}{2} \partial^2_{x_t^i}- 2\sum_{l\neq i=1}^m \left( \frac{h_l}{\left(x^l - x^i\right)^2} - \frac{1}{\left( x^l - x^i\right)} \partial_{x^l}
\right) \nonumber\\
&\quad&\!\!\!\! - 2\sum_{k=1}^n \left( \frac{h_k}{\left(z^k - x^i\right)^2} - \frac{1}{\left( z^k - x^i\right)} \partial_{z^k}
\right) + a.h.\nonumber\\
\end{eqnarray}
where $a.h.$ denotes the antiholomorphic contributions which, in the following, we will drop for simplicity.
Additionally, we do not want to spend time on the case that two fields fuse to the identity, hence we will leave that part out in the following considerations, too.

In a situation with $x^i \rightarrow z^j,$ (i.\,e.\ the {\sc L\"owner} differential equation drives the $j^{\text{th}}$ field at $z_t^j$ to the real axis where the tip of the $i^{\text{th}}$ {\sc SLE} trace is growing), we have two effects if $h_j = h_i \in \lbrace h_{(1,2)}, h_{(2,1)}\rbrace $: 
For two degenerate fields corresponding to the representation $\phi_{(r,s)}$ of the same weight on level $r\cdot s = 2$ in the {\sc Ka\u{c}} table, we know that the outcome of fusing them is (apart from the identity) a degenerate field corresponding to $\phi_{(r',s')}$ with $r' = 2r -1$ and $s' = 2s -1$ on level $r'\cdot s' = 3$. When acting on the appropriate correlation function\ldots:
\begin{itemize}
 \item[(a)] the $i$-th differential operator changes ($z^j - x^i = \epsilon$, $2z = z^j + x^i$) : 
\begin{equation}\label{eq:bulkboundarylimitop}
\mathcal{D}_{-2}^{m,n}(x^i;\{x^l\},\{z^k\}) \rightarrow 
\epsilon \mathcal{D}_{-3}^{m,n-1}(z;\{x^l\},\{(z^k,\bar{z}^k)\}_{k\neq j})  
\end{equation}
where $\mathcal{D}_{-3}^{m,n-1}(z;\{x^l\},\{(z^k,\bar{z}^k)\}_{k\neq j})$ is the differential operator imposing the null vector differential equation on a correlation function including a primary field of weight $h=h_{(r',s')}$:
\begin{eqnarray}\label{eq:D3bulkboundaryop}
&\mathcal{D}_{-3}^{m,n-1}&\!\!\!\!(z;\{x^l\},\{(z^k,\bar{z}^k)\}_{k\neq j}) \\
&:\,=&\!\!\!\!\!\!\!\!\frac{\kappa}{2} \partial^3_{x^i} - 4\sum_{l\neq i=1}^m \left( \frac{h_l}{\left(x^l - x^i\right)^2} - \frac{1}{\left( x^l - x^i\right)} \partial_{x^l}
\right) \partial_{x^i}\nonumber\\
&\quad&\!\!\!\!\!\!\!\!-\, 4\sum_{k\neq j=1}^n \left( \frac{h_k}{\left(z^k - x^i\right)^2} - \frac{1}{\left( z^k - x^i\right)} \partial_{z^k}
\right)\partial_{x^i} + a.h.\nonumber\\
&\quad&\!\!\!\!\!\!\!\!+ \,4\left(\frac{8}{\kappa}-1\right)\sum_{l\neq i=1}^m \left( \frac{2h_l}{\left(x^l - x^i\right)^3} - \frac{1}{\left( x^l - x^i\right)^2} \partial_{x^l}
\right) \nonumber\\
&\quad&\!\!\!\!\!\!\!\!+ \,4\left(\frac{8}{\kappa}-1\right)\sum_{k\neq j=1}^n \left( \frac{2h_k}{\left(z^k - x^i\right)^3} - \frac{1}{\left( z^k - x^i\right)^2} \partial_{z^k}\partial_{x^i}
\right) + a.h.\,,\nonumber
\end{eqnarray}
where $a.h.$ denotes the antiholomorphic contributions.
 \item[(b)] for $\epsilon \rightarrow 0$, the weight $h_j$ changes to $h_{(r',s')}$ in all of the other $m-1$ equations.
\end{itemize}
In the next part we will repeat the same considerations for the arch configuration observable. Of course, we could also investigate the situation where $x_t^i \rightarrow x_t^{k}$. However, all interesting features of that case will be contained in the one discussed in the next part so that we did not want to include it here. Additionally, $z_t^j \rightarrow z_t^l$ will not be discusses since it would only lead to a different observable and no new differential equations or changes of the {\sc SLE} quantities.


\subsection{The Purely Boundary Differential Operators}
In multiple {\sc SLE} as investigated in \cite{Graham:2005}, we do not only get the differential operators (\ref{eq:bulkboundaryop}) but also others relating only the boundary fields. They show up when we consider the observable that tells us which arch configuration is present in the theory. It corresponds to the correlator of the boundary {\sc CFT} fields \cite{Bauer:2005jt} and hence vanishes under the action of:
\begin{equation}\label{eq:boundaryboundaryop}
\mathcal{D}_{-2}^m(x^i;\{x^l\}_{l\neq i}) := \frac{\kappa}{2} \partial^2_{x^i} -2\sum_{l\neq i=1}^m \left( \frac{h_l}{\left(x^l - x^i\right)^2} - \frac{1}{\left( x^l - x^i\right)} \partial_{x^l}
\right)\,.
\end{equation}
To investigate the different arch configurations that can show up in a theory with $m$ traces starting at the boundary, it is natural to look at the effects when $x^i \rightarrow x^{i+1}$. Obviously, the traces could either pair up or just come close to each other without touching. The first situation has already been identified with fusion to the identity \cite{Bauer:2005jt} so that we do not want to consider that part here. Note that only neighboring {\sc SLE} traces are able to fuse when choosing $a_t^k =1$ \cite{Graham:2005}; in any case other situations might lead to intersection problems of the {\sc SLE} traces so that we do not want to consider them here. 

Similarly to the situation of bulk-boundary fusion, in this limit the differential operator $\mathcal{D}_{-2}^m(x^i)$ becomes dependent on $\delta := x^{i+1} -x^i$ and $x:= (x^i+x^{i+1})/2$. However, the situation is slightly different since the differential operators $\mathcal{D}_{-2}^m (x^k)$ for $k\neq i,i+1$ also depend on $\delta$ and $x$, so that we have to change all differential operators. Fortunately, this is an easy work which can be solved by a simple expansion in the small parameter $\delta$. Thus we get (again dropping the part where the fields fuse into the identity) when acting on the appropriate correlation function:
\begin{equation}
\mathcal{D}_{-2}^m(x^i;\{x^l\}_{l\neq i}) \rightarrow 
\delta \mathcal{D}_{-3}^{m-1}(x;\{x^l\}_{l\neq i,i+1})\,,  
\end{equation}
where $\mathcal{D}_{-3}^{m-1}(x;\{x^l\}_{l\neq i,i+1})$ is analogously defined to (\ref{eq:D3bulkboundaryop}) and for $j\neq i, i+1$:
\begin{equation}
\mathcal{D}_{-2}^m(x^j;\{x^l\}_{l\neq j}) \rightarrow 
\mathcal{D}_{-2}^{m-1}(x^j;x,\{x^l\}_{l\neq j,i,i+1}) \,,
\end{equation}
with $h = h_{(r',s')}$ as introduced in the previous section showing up as the weight of the field depending on $x$.

As said above, the case where two {\sc SLE} paths not only come close to each other but acutally meet at $t_0$ has already been discussed in \cite{Bauer:2005jt,Graham:2005}. This situation corresponds to a kind of annihilation process of two {\sc SLE} tips; the traces vanish from the upper half plane for $t>t_0$, hence being identified with the fusion branch that results in the identity field. However, it is quite more interesting to study the other part of the fusion process -- although the probability of such an event gets small for small distances, the question which power law can be assigned to it may reveal the correct correspondence to {\sc bCFT} quantities. 

Together with the results in section \ref{sec:Obs}, 
we will show in the following how we can relate a boundary field approaching a field of an observable $\mathcal{O}$ or another boundary field to the probability of intersecting a ball of radius $r \asymp \epsilon$. Note that these events in the {\sc SLE} picture are an {\sc SLE} trace approaching a point in the upper half plane or two {\sc SLE} traces approaching each other, respectively.
\section{Interpretation of Merging multiple SLE traces}
In this section we will try to relate the results of the previous two sections via identifying the scaling behavior of the {\sc SLE} probability $\textbf{P}_{\epsilon,z_0}$ and that of correlation functions after the fusion of fields in {\sc bCFT}. 

\subsection{SLE Traces Visiting a Point in the Upper Half Plane}\label{sec:SLEpoint}
Let us consider the following situation: we have a multiple {\sc SLE} with $m$ interfaces $\gamma^k$ starting from the boundary and going up to $\infty$. Additionally, we have an observable given by the correlation function of $n$ primary fields, located at the points $z_0^l$ in the upper half plane: $O(\lbrace z_0^l,\bar{z}_0^l\rbrace) = \prod_{l=1}^n \phi_l(z_0^l,\bar{z}_0^l)$. We will only consider the case where one of them, say the $j^{th}$ has the same weight as the {\sc bCFT} field corresponding to  the $i^{th}$ interface -- $h_j = h_i = h_{(r,s)}$ with $(r,s)  \in \lbrace (1,2),(2,1) \rbrace$. For this situation, we have $m$ {\sc SLE} martingales as shown in section \ref{sec:Obs}.

Going back to what we learned before, we take a look at the differential operators given by the {\sc L\"owner} equation (\ref{eq:bulkboundaryop}) and their limits for $x^i \rightarrow z^j$ (\ref{eq:bulkboundarylimitop}):
\begin{equation}
\mathcal{D}_{-2}^{m,n}(x^i;\{x^l\},\{z^k\}) \rightarrow 
\epsilon \mathcal{D}_{-3}^{m,n-1}(z;\{x^l\},\{(z^k,\bar{z}^k)\}_{k\neq j})  
\end{equation}
leaving out the identity part as said above. 

Additionally, we remember that the {\sc SLE} martingale $O^{\text{\sc SLE}}_t(x_t^i,\{x_t^l\},\{z_t^k\})$ can be expressed via the {\sc CFT} expectation value of our observable $O$
\begin{equation}
\mathcal{O}^{\text{\sc CFT}}(x^i,\{x^l\},\{z^k\}) :\,=\frac{\left\langle \psi (\infty),\psi_{(r,s)}(x^i) \phi_{(r,s)}(z^j)\prod\limits_{k\neq j=1}^n \phi (z^l) \prod\limits_{l\neq i = 1}^m \psi (x^l) \right\rangle}{\left\langle \psi (\infty),\psi_{(r,s)}(x^i) \phi_{(r,s)}(z^j)\prod\limits_{l\neq i = 1}^m \psi (x^l) \right\rangle}\,.
\end{equation}
Inserting the {\sc OPE} for $\psi_{(r,s)}(x^i) \phi_{(r,s)}(z^j)$,
the differential equation 
\begin{equation}
 \mathcal{D}^{m,n}_{-2} \mathcal{O}^{\text{\sc CFT}}(x^i,\{x^l\},\{z^k\}) = 0
\end{equation}
shows the following behavior as $\epsilon :\,= x^i - z^j \rightarrow 0$:
\begin{equation}
\epsilon \mathcal{D}^{m,n-1}_{-3}\epsilon^{-\mu} 
\tilde{\mathcal{O}}^{\text{\sc CFT}}(x,\{x^l\}_{l\neq i},\{z^k\}_{k\neq j})
=0\,,
\end{equation}
introducing $-\mu = h_{(r',s')}-2h_{(r,s)}$, $2x = x^i+z^j$ and $\tilde{\mathcal{O}}^{\text{\sc CFT}}$ as the expectation value with $\psi_{(r,s)}(x^i) \phi_{(r,s)}(z^j)$ fused to $\psi_{(r',s')}(x)$.


Remembering equation (\ref{eq:expectation}), i.\,e.\ 
\begin{equation}
\frac{\text{d}}{\text{d}t} \mathbb{E} (\mathcal{O}_t^{\text{\sc SLE}}) = \mathcal{D}_{-2}^{m,n}\mathcal{O}^{\text{\sc CFT}}\left(x^i,\{x^l\},\lbrace (z^k,z^*{}^k)\rbrace\right)\,,
\end{equation}
we have just shown that the {\sc r.h.s} behaves like
\begin{equation}
\epsilon^{1-\mu}  
\end{equation}
for small values of $\epsilon$. Note that this should be interpreted as a functional relation: since the dependency on the distance $\epsilon$ is the same for all times, as viewn as a functional dependency it is time independent in some sense.


The outcome on the {\sc l.h.s.} is more easily to find: The expectation value of a martingale conditioned on a certain event is given by the expectation value of that specific martingale times the probability of the event. Obviously, in this case the event is that of an {\sc SLE} trace coming at least $\epsilon$ close to the point $z_j$ in the upper half plane. Hence we have: 
\begin{equation}
\frac{\text{d}}{\text{d}t} \mathbb{E}(\mathcal{O}_t^{\text{\sc SLE}}) \rightarrow  \frac{\text{d}}{\text{d}t}  \textbf{P}_{\epsilon,z_j}\mathbb{E}(\tilde{\mathcal{O}}_t^{\text{\sc SLE}})\,.
\end{equation}
Now we will again assume, that the {\sc SLE} expectation value of the martingale corresponds via our statistical physics picture to the respective correlation function in {\sc bCFT}. Hence we expect the scaling behavior of $\textbf{P}_{\epsilon,z_j}$ to be the same as that of the {\sc r.h.s.}. Remembering from section \ref{sec:SLEprob} that the probability of an {\sc SLE} trace intersecting a ball of radius $\epsilon$ located at $z_j$ in the upper half plane sufficiently far away from the boundary:
\begin{equation}
\textbf{P}_{\epsilon,z_j}  \asymp  \epsilon^{2-d_{K}} \,.
\end{equation}
In order to see that $1-\mu$ is indeed equal to $2-d_K$, we have to insert some knowledge about the weights of the primary fields and the dimension of the {\sc SLE} hull. Via the formula for the central charge we can deduce that
\begin{equation}
 h_{(r,s)} = \frac{(r\kappa -4s)^2 - (\kappa-4)^2}{16 \kappa}
\end{equation}
which means that 
\begin{equation}
\begin{array}{llcc}
 h_{(1,2)} = \frac{6-\kappa}{2\kappa} & h_{(1,3)} = \frac{8-\kappa}{\kappa} &\rightarrow &\mu = \frac{2}{\kappa} \,,\\
 h_{(2,1)} = \frac{3\kappa-8}{16} & h_{(3,1)} = \frac{\kappa-2}{2} &\rightarrow &\mu = \frac{\kappa}{8}\,.
\end{array}
\end{equation}
Hence we have $1-\mu \in \{1-\frac{\kappa}{2},1-\frac{8}{\kappa}\}$ which is the same as $2-d_K$ as $d_K \in \{1+\frac{\kappa}{2},1+\frac{8}{\kappa}\}$.

Therefore the event of the outer hull of the {\sc SLE} coming close to a point can be identified\symbolfootnote[3]{Of course, here ``identifying'' is not meant in a mathematically rigorous sense!} with fusion of the {\sc bCFT} boundary field with a bulk field of the same dimension.


Of course, the other $m-1$ martingales $M_t^k$ and correlation functions $M^k$, $k \neq i$ are also affected as discussed in section \ref{sec:scaling1}. However, this does not lead to any interesting behavior since the only changing quantity is the observable, but at a point sufficiently far away from any of the other $m-1$ {\sc SLE} traces and $n-1$ points with primary fields attached.
\subsection{SLE Traces Merging}\label{sec:SLEtrace}
Certainly, the point $z_j$ we have been looking at could also be lying on one of the other {\sc SLE} traces. Hence, in complete analogy to the considerations above, we will treat the situation of two {\sc SLE} traces approaching each other. Obviously, this corresponds to a specific arch configuration. Thus we take the results from section \ref{sec:probBCFT}, where we investigated the correlation function of boundary fields 
\begin{equation} 
\langle\psi_{\alpha_\infty,\alpha_1}( \infty),  \psi_{\alpha_1,\alpha_2}(x^1) \ldots \psi_{\alpha_m,\alpha_\infty}(x^m) \rangle\,.
\end{equation}
As we already argued, in {\sc bCFT} this is the probability $\textbf{P}_{\gamma_t^1,\ldots,\gamma_t^m}$ that this special configuration emerges by chance. Additionally, as a byproduct of the calculations including an observable, it has been proven in \cite{Bauer:2005jt} to be a martingale, too. As the boundary changing operators are taken to be primary fields with weight $h_i=h_{i+1} = h_{(r,s)}$ with $(r,s) \in \lbrace (1,2),(2,1)\rbrace$, the correlator vanishes if we let the operators $\mathcal{D}_{-2}^m(x^i,\{x^k\}_{k\neq i})$ act on it.

In the limit $x^i \rightarrow x^{i+1}$, we know that the {\sc CFT} quantity
\begin{equation}
\mathcal{D}_{-2}^m(x^i,\{x^k\}_{k\neq i}) \langle \ldots \psi_{\alpha_i,\alpha_{i+1}}(x^i)\psi_{\alpha_{i+1},\alpha_{i+2}}(x^{i+1})\ldots \rangle
\end{equation}
splits up into two parts of which the identity part will be left out -- as already stated above, it is uninteresting since the situation becomes essentially the same as for $m-2$ {\sc SLE}s starting at a later point in time \cite{Bauer:2005jt}. The other part is proportional to
\begin{equation}
\mathcal{D}_{-3}^{m-1}(x,\{x^k\}_{k\neq i,i+1})\delta^{1-\mu}\langle \ldots \psi_{\alpha_i,\alpha_{i+2}}(x)\ldots \rangle \,.
\end{equation}
with $1-\mu = 2-\text{d}_{K}$ as already shown in the previous section. Hence we can deduce that the probability that two {\sc bCFT} fields $\psi_{(r,s)}$ are fusing to $\psi_{(r',s')}$ can be expressed as the corresponding {\sc SLE} traces coming close to each other, since 
if we look at the upper half plane at time $t$ before applying $G_t$, we see that if the $i^{\text{th}}$ {\sc SLE} trace approaches the $(i+1)^{\text{st}}$ near the point $z_0 \in \{ G_t^{-1}(x_t^{i+1})\}$ \symbolfootnote[2]{Of course $z_0 \notin \mathbb{H}/K_t$, hence this formula should rather be interpreted as the limit $z_0^{i+1} \in \{ \lim_{\epsilon \rightarrow 0} G_t^{-1}(x_t^{i+1}+ \mathrm{i} \epsilon )\}$.}, the {\sc SLE} probability is given by 
\begin{equation}
 \textbf{P}_{\delta,z_0^{i+1}} \propto \delta^{2-\text{d}_{K}} = \delta^{1-\mu}  
\,,
\end{equation}
thus exhibiting the same functional $\delta$ dependence as $ \mathcal{D}^{m-1}_{-3} \textbf{P}_{\gamma^1,\gamma^{i-1},\gamma^{\text{merged}}\gamma^{i+2},\gamma^m} $. 
Thus fusion of boundary fields in {\sc CFT} can be considered as merging of {\sc SLE} hulls. More precisely: merging {\sc SLE} traces of speed\ldots
\begin{itemize}
 \item[(i)] \ldots$4< \kappa < 8$ corresponds to fusing two $h_{(1,2)}$ to an $h_{(1,3)}$ boundary field, 
 \item[(ii)] \ldots$0 < \kappa < 4$ corresponds to fusing two $h_{(2,1)}$ to an $h_{(3,1)}$ boundary field.
\end{itemize}
\subsection{The Angular Dependency}\label{sec:SLEang}
Unfortunately, we are not able to give a complete satisfactory interpretation to the angular dependency of $\textbf{P}_{\epsilon,z_0}$. Hence this section will be quite sketchy but we are convinced that the basic recipie how to achieve this goal works in principle. Our proposal is based on the connection of the absolute value $|z_0|$, the imaginary part $\mathfrak{Im}(z_0)$ and the angle $\alpha(z_0)$ between the real axis and the vector pointing to a point $z_0$:
\begin{equation}\label{eq:absang}
 |z_0| = \mathfrak{Im}(z_0) \cdot \sin \alpha(z_0)^{-1}\,,
\end{equation}
as has been attempted in \cite{Bauer:2004}. Thus we will give a short handwavy description on our thoughts on this subject.

Equation (\ref{eq:absang}) could be used after applying three known properties of $n$-point functions:
\begin{itemize}
 \item[(i)] behavior under conformal transformations for a fixed time $t$:
\begin{eqnarray}
 \langle \ldots \psi(x_t) \ldots \rangle &=& \langle \ldots \psi(G_t(z_0))\ldots \rangle \nonumber\\
&=& (G_t'(z_0))^{-h} \langle \ldots \psi(z_0)\ldots \rangle \,,\label{eq:G_tbound}\\
 \langle \ldots \phi(z_t,z_t^*) \ldots \rangle &=& \langle \ldots \phi(G_t(z_t),G_t^*(z_t^*))\ldots \rangle\nonumber\\ 
&=& |G_t'(z_0)|^{-2h} \langle \ldots \phi(z_0,z_0^*)\ldots \rangle \,,\label{eq:G_tbulk}
\end{eqnarray}
for local bulk fields $\phi(z_0,z_0^*)$ with weights $h_0 = \bar{h}_0 =h$ and boundary fields $\psi(x_t)$ of weight $h$. Note that the interpretation of (\ref{eq:G_tbound}), (\ref{eq:G_tbulk}) is difficult in the standard {\sc CFT} context since we only have the formalism to deal with the whole upper half plane -- not $\mathbb{H}/K_t$. Additionally, it is not clear what $G_t$ might be in the limit of two traces coming close to each other. These equations should therefore just be used to illustrate how the angular dependence on $\alpha(z_0)$ could in principle be recovered from the general correlation function.
 \item[(ii)] cluster decomposition: if we separate the variables of a correlation function into two sets and scale one set towards $\infty$, the behavior of the correlation function is dominated by the product of two correlation functions, corresponding to the two sets of variables, multiplied by an appropriate power of the separation \cite{Gaberdiel:2000}:
\begin{equation}
 \left\langle \prod_i \phi(\lambda w_i) \prod_j \phi(z_j) \right\rangle \sim  \left\langle \prod_i \phi(w_i) \right\rangle \left \langle \prod_j \phi(z_j) \right\rangle \lambda^{-\sum_i h_i}\,.
\end{equation}
For our purposes, we will separate the two fusing fields from the rest which leaves us with a two point function of two $h_{(r,s)}$ fields with $(r,s) \in \lbrace(1,2),(2,1)\rbrace$ which is nonzero and the correlation function of the rest. Afterwards we insert the {\sc OPE}, choose the non-identity part including the $h_{(r',s')}$ field with $r'=2r-1$ and $s' = 2s-1$ as well as its descendants. Here it is of utmost importance that we are dealing with {\sc bCFT} where the one point functions of boundary fields does not vanish.

In the theory of probability, this corresponds to independent events $X,Y$:
\begin{equation}
\mathbb{E}(XY) = \mathbb{E}(X)\cdot \mathbb{E}(Y) \,.
\end{equation}
This assumption is valid if the other {\sc SLE} traces and the rest of the observable is far away from the interesting point.
 \item[(iii)] conformal invariance then leaves us with the following prefactor:
\begin{equation}\label{eq:prefactor}
 \epsilon^{-\mu}|z|^{-h_{(r',s')}}\,.
\end{equation}
where $\epsilon$ is the distance between the two fusing fields and $z$ their mean coordinate.
\end{itemize}
Note that for the bulk field case, we have to compare it to the product of two probabilities since this implies that the mirrored {\sc L\"owner} trace approaches the mirrored point $z_0^*$ -- hence we are looking for a correspondence to $\textbf{P}_{\epsilon,z_0}^2$.

The behavior near the boundary is only important in the case of $4< \kappa <8$ where we have $h_{(1,3)}$ type fields since otherwise the {\sc SLE} trace is simple. Thus for $t>0$ it neither touches the boundary nor itself and therefore we can assume that it does not come near the boundary again. Hence, we have $h_0 = h_{(1,3)} = (8-\kappa)/\kappa$
and the factor $\sin \alpha(z_0)^{8/\kappa-1}$ shows up squared for $h_0 = \bar{h}_0$ in $\textbf{P}_{\epsilon ,z_0}^2$ and to power one in $\textbf{P}_{\delta,x_0}$. Note that our sine exponent differs from the one mentioned in \cite{Bauer:2004} but is indeed the same as given in \cite{Beffara:2002} (cf.\ (\ref{eq:Pepsz0})).


Nevertheless, the situation is not that easy to be solved since -- apart from the problems already listed above -- it is still an open problem how to translate the probability $\textbf{P}_{\epsilon,z_0}$ to the multiple {\sc SLE} properly. It has been derived for a situation where a single {\sc SLE} starts at $0$, exhibiting an angle $\alpha(z_0)$ which is measured at the origin. Hence what has to be done is to extend the formalism to any reference point $\xi_t$ which should be an easy problem using translation invariance of the problem. As a second step, more than just one {\sc SLE} trace should be considered which will surely lead to modifications, too, and probably requires a definition of multiple {\sc SLE} on the disk. However, this would go beyond the scope of our paper which is dedicated to the {\sc SLE}-{\sc bCFT} correspondence and not intended to focus on new {\sc SLE} results. Hence we just argue that if we assume that all other {\sc SLE} traces and observables are sufficiently far away from the $z_0$, we can apply the cluster decomposition theorem for local quantum field theories. 
Hence, equation (\ref{eq:prefactor}) gives us an independent prefactor of a correlation function of the rest of the other fields involved in the theory. This way, the asymptotic behavior stays the same and our considerations should be valid in general $m$ multiple {\sc SLE} with any observable $O$ given by a product of local primary fields.

\section{Multiple SLE as Martingale Weighted SLE}
Following the idea in \cite{Werner:2004} for {\sc SLE}($\kappa,\vec{\rho}$), we will argue that multiple {\sc SLE} can be interpreted as $m$ single {\sc SLE}s moving in the same upper half plane weighted by the martingale $M_t$ given in (\ref{eq:martingale}). Additionally, we give a probability interpretation of this martingale following \cite{Dubedat:2004-2,Dubedat:2005} and discuss the case of collapsing starting points. 
\subsection{The Associated Martingale $M_t$}\label{sec:radnik}
Suppose we have a random process $(X_t)$, $t \geq 0$, taking values in some space $X$. Now let us define a function $f\,:\,X\rightarrow [0,\infty)$ such that $f(X_t)$ is measurable with bounded expectation value. In that case, we are able to weight the probability measure $P_{\text{old}}$ by $f(X_t)$ and hence consider a new probability measure $P_{\text{new}}$ which can be achieved by the {\sc Radon-Nikodym} derivative:
\begin{equation}
 \frac{\text{d} P_{\text{new}}}{\text{d} P_{\text{old}}} = \frac{f(X_t)}{\mathbb{E}(f(X_t))}\,.
\end{equation}
In our case, the new law of $X_t$ under the new measure can be determined explicitly. Let $\mathcal{F}_t$ be the filtration to which $X_t$ is adapted. Then $M_t :\,= \mathbb{E}( f(X_t) | \mathcal{F}_t)$ is a martingale by construction and for every $A \in \mathbb{F}_t$ we have $P_{\text{new}}(A) = \mathbb{E}(1_A M_t)/M_0$. With the help of {\sc Girsanov}'s theorem, we can compute the change of the drift term due to the weighting with the martingale $M_t$ explicitly. For any local martingale $N_t^{\text{old}}$ under the old measure, we get a new local martingale $N_t^{\text{new}}$ via:
\begin{equation}
 N_t^{\text{new}} = N_t^{\text{old}} - \int_0^t \frac{\text{d} \langle N^{\text{old}},M \rangle_{t'}}{M_{t'}}\,.
\end{equation}
In our case, it has been derived in \cite{Graham:2005} that the weight martingale $M_t$ is given by (\ref{eq:martingale}) and thus leads to the change in the drift terms stated in (\ref{eq:driftchange}).

Hence we can say that multiple {\sc SLE} is just ordinary {\sc SLE} under a new measure.
\subsection{Probability Interpretation for $M_t$}\label{sec:probInt}
Analogously to the ideas presented in \cite{Dubedat:2004-2,Dubedat:2005,Dubedat:2006}, we will give a probability interpretation to the martingale found for multiple {\sc SLE} processes ( \ref{eq:martingale}). However, we have to point out the crucial difference between our ansatz and the considerations made in \cite{Dubedat:2004-2,Dubedat:2005}: in our opinion, multiple {\sc SLE} does not correspond to {\sc SLE}($\kappa,\vec{\rho}$) with $\vec{\rho} = (2,2,\ldots,2)$. In terms of the formalisms typically used for the description of {\sc SLE}($\kappa,\vec{\rho}$), the stochastic nature of the coordinates of the force points is only due to the {\sc Brownian} motion of the {\sc SLE} coordinate although for $\rho_i = 2$ the $i^{\text{th}}$ field would be of the correct weight. However, in multiple {\sc SLE}, every field that corresponds to an interface exhibits its own {\sc Brownian} motion. 
However, if we take {\sc SLE}($\kappa,\vec{\rho}$) as a special case of multiple {\sc SLE} with $\kappa_i = a_t^i = 0$ for all $i$ except one and the partition function to be that of the {\sc Coulomb} gas formalism as described in \cite{Cardy:2004xs,Kytoelae:2006}, the two {\sc SLE} types become the same. We will come back to that interpretation in a forthcoming paper \cite{Mueller-Lohmann:2008-1}.

For $h_i = \frac{6-\kappa_i}{2\kappa_i} = h_{(1,2)}  \geq \frac{5}{8}$ , i.\,e.\ $\kappa_i \leq \frac{8}{3}$, let us consider the single {\sc SLE} picture with $g_t^i$ as the $i^{\text{th}}$ of $m$ {\sc L\"owner} mappings of a multiple {\sc SLE} setting with driving function $w_t^i = \sqrt{\kappa}_i B_t^i$ as in (\ref{eq:singleSLE}). For $i=1,\ldots,m$, let $\mathfrak{O}_{c_i}$ be independent realizations of the {\sc Brownian} loop soup 
of intensity $-c_i$. 
Let $K_i$ be the $i^{\text{th}}$ hull generated by $\gamma_i [0,\infty ) $
and the loops that intersect it. For each $i$ we know from the properties of the {\sc Brownian} loop soup that the probability that $K_i$ intersects some hull $A_i$ is
\begin{equation}\label{eq:probBubble}
\textbf{P}_i:\,= \textbf{P}\lbrace K_i \cap A_i = \emptyset \rbrace = \Phi_{A_i}^{'} {}^{h_i} \exp \left( -\frac{c_i}{6} \int_0^t \mathcal{S} \Phi_s(w_s^i) \text{d}s\right)\,,
\end{equation}
where $\Phi_{A_i}: \mathbb{H} / A_i \rightarrow \mathbb{H}$ and, again, $\mathcal{S}$ the {\sc Schwarzian} derivative.

Now let us consider multiple {\sc SLE} with the multiple {\sc L\"owner} mapping as stated in (\ref{eq:multipleLoewner}).
Let $A_i$ be the hull created by $\bigcup_{j \neq i}K_j$. Then $H_t^i(w_t^i)=x_t^i$ as defined in section \ref{sec:multSLE} plays the role of $\Phi_{A_i}$ in this context.
Following the argumentation in \cite{Lawler} 
we define 
\begin{eqnarray}
M_t&:\,=&\frac{Z[x_t]}{Z[x_0]} \prod_{i=1}^m 
\left(H_t^i{}'(w_t^i)\right)^{h_i} \exp \left( -\frac{c_i}{6} \int_0^t \mathcal{S} H_s^i(w_s^i) c_s^i \text{d} s \right) \nonumber\\
&\quad& \quad \quad\quad\quad \exp \left( -\int_0^t \frac{1}{Z[x_s]} \mathcal{D}_{-2}^m(x_s^i) Z[x_s] a_s^i \text{d} s\right)
\end{eqnarray}
on $\lbrace t < t_A \rbrace$ with $t_A$ being the intersection time. $M_t$ is a martingale and $0 \leq M_t\leq \left(H_t^i{}'(w_t^i)\right)^{h_i} \leq H_t^i{}'(w_t^i) \leq 1$ for $\alpha_i \geq \frac{5}{8}$ as shown in \cite{Graham:2005}. 

Knowing that the level two null vector differential equation imposed by the action of $\mathcal{D}_{-2}^m(x_s^i)$ is satisfied since $h_i = h_{(1,2)}$, we can drop the last factor. 
Taking a look at the expectation value of $M_t$ and using (\ref{eq:probBubble}), it is easy to see that
\begin{equation}
 \mathbb{E} (M_t) = \mathbb{E} \left( \frac{Z_[x_t]}{Z[x_0]} \right) \prod_{i=1}^m\textbf{P}_i\,.
\end{equation}
As already argued in section \ref{sec:probBCFT}, the ratio of the partition functions just gives us the probability, that we have an arch configuration starting at arbitrary points without traces pairing up. Hence, the expectation value of our martingale $M_t$ equals the probability $\textbf{P}^{m\gamma}$ that we have $m$ traces that do neither intersect each other nor pair up to form archs.


\subsection{Fusion in the Associated Martingale}\label{sec:fusmart}
Now we will consider two points collapsing, i.\,e.\ $x_t^{i+1}\rightarrow x_t^i$, with $h_i = h_{i+1} = h_{(1,2)}$. Note that we still restrict ourselves to the case where the tips are not meeting at their endpoints to pair up. Using $x_t^k = H_t^k(w_t^k)$, we get for the first factor:
\begin{eqnarray}
\lim_{x_t^i\rightarrow x_t^{i+1}}\frac{Z[x_t]}{Z[x_0]}
&=&  \lim_{w_t^i\rightarrow w_t^{i+1}}  
\left(H_t^{i+1}(w_t^{i+1})-H_t^i(w_t^i)\right)^{-\mu}\frac{\tilde{Z}[x_t]}{Z[x_0]}\nonumber\\
&\propto& 
\left(H_t^{'}(w)\right)^{-\mu}\frac{\tilde{Z}[x_t]}{Z[x_0]}
\end{eqnarray}
with $\mu = h_i +h_{i+1} - h$ and $\tilde{Z}$ denoting the partition function wherein we fused the $i^{\text{th}}$ and $(i+1)^{\text{st}}$ boundary field of weight $h$ at coordinate $w_t = (w_t^i + w_t^{i+1})/2$. From {\sc CFT} fusion rules we know that $h=h_{(1,1)}$ or $h=h_{(1,3)}$. As we agreed to restrict ourselves to the second case (see above), we have in this limit
\begin{equation}
\textbf{P}^{m\gamma}_{x_t^i \rightarrow x_t^{i+1}} \propto \left(\phi_{A_i}'\right)^{h_{(1,3)}} \prod_{j\neq i,i+1}\left(\phi_{A_j}'\right)^{h_j} \,,
\end{equation}
switching back to the notation used in (\ref{eq:probBubble}).

This can be easily extended to the case of $m \geq 2$ curves and hence in the limit of all points collapsing to $x_t$, we get:
\begin{equation}
\textbf{P}^{m\gamma}_{\lbrace x_t^k\rbrace \rightarrow x_t} \rightarrow \left(\phi_{A_i}'\right)^{h_{(1,m+1)}}\,.
\end{equation}
in complete agreement with {\sc CFT} fusion rules. Hence we will take this as another hint at the picture of {\sc SLE} tips corresponding to boundary {\sc CFT} fields of weight $h_{(1,2)}$. Note that this behavior has also been observed in \cite{Dubedat:2004-2,Dubedat:2006} for a different picture of multiple {\sc SLE}; a similar case is studied for {\sc SLE}($\kappa,\vec{\rho}$) in \cite{Dubedat:2005}.
Of course, we should not be suprised that this result is similar to that of {\sc SLE}($\kappa,\vec{\rho}$) as the latter is only a special case of multiple {\sc SLE} which we will show in an forthcoming paper \cite{Mueller-Lohmann:2008-1}. 

Moreover, if we let $h_i = \frac{3\kappa_i-8}{16}\geq \frac{5}{8}$, i.\,e.\ $\kappa_i \rightarrow 16/\kappa_i \leq \frac{8}{3}$ which means that $\kappa_i \leq 6$ and $c_i \leq 0$, we cover the $h_i = h_{(2,1)}$ fusion to $h_{(m+1,1)}$ case. 

Additionally, in complete agreement to the results of the appendix of \cite{Graham:2005}, we could also set $a_t^i = 1 \quad \forall i$. It has been shown that this corresponds to a setting without mutual intersections of the {\sc SLE} traces. Inserting that in the notation used in \cite{Graham:2005}, $c_t^i =1$, we have $\left(H_t^i{}'(w_t^i)\right)^2 = 1$ and hence  $\textbf{P}_i = 1$ which also means nothing else than no intersections occur w.p.1.

\section{Discussion}
In this paper we have shown how to interpret merging {\sc SLE} traces and traces approaching a marked point $z_j$ as {\sc CFT} quantities via their scaling behavior. The outcome is quite natural: locally, these events can be interpreted as the fusion of two 
 $h_{(r,s)}$ to an $h_{(r',s')}$ field with $(r,s) \in \{ (1,2),(2,1)\}$ and $r' = 2r-1$, $s' = 2s-1$. 

However, this interpretation is only valid in a small region around $z_j$ due the nature of the{\sc SLE} mapping. It erases any event from the scene so that for later times, the event is gone and leaves no hint at all that it ever happened. 

The two events -- merging traces and traces approaching a marked point -- may be distinguished by also taking the antichiral part of the bulk field into account. We have proposed a way how this can be done by considering the {\sc SLE} probability of visiting a point near the boundary where the initial behavior takes over in terms of an angular dependency. In conctrast to previous papers \cite{Bauer:2004}, we are able to give an interpretation of the sine exponent $h_{(1,3)}$ initially derived in \cite{Beffara:2002} and have been able to give a hint how to extend the interpretation to the general case of $n$ point functions. 

Interestingly, the assignment of the different {\sc SLE} phases -- $0 < \kappa < 4 $ and $4 < \kappa < 8$ -- to different types of fields -- $h_{(3,1)} = \frac{\kappa-2}{2}$ and $h_{(1,3)} = \frac{8-\kappa}{\kappa}$ -- has already been observed in another context by Riva and Cardy \cite{Cardy:2006} while investigating the $Q$-states {\sc Potts} model and its relation to {\sc SLE}.

In order to achieve our results, we have tried a novel approach to the view on the {\sc SLE}-{\sc bCFT} correspondence: in our opinion, the mapping between the two theories goes via the identification of whole differential equations. Hence we can not close our eyes upon the behavior of the differential operators. This inevitably leads us to additional exponents that enable us to give a quite natural interpretation  while staying completely within the {\sc Ka\u{c}} table of minimal {\sc bCFT} models for candidates for interpretation.

Additionally we picked up the viewpoint of all {\sc SLE}s being just ordinary {\sc SLE}s under different measures to give a probability interpretation to the weighting martingale. We pointed out how, again, other {\sc Ka\u{c}} table weights show up as scaling exponents in multiple {\sc SLE} settings in complete agreement with what we would expect from {\sc CFT} fusion rules. 
\subsection*{Acknowledgements}
I would like to thank Greg Lawler for helpful comments at the IPAM Random Shapes Workshop at UCLA this spring. Additionally I would like to express my gratitude to Kalle Kyt\"ol\"a for beneficial discussions and Stas Smirnov, Scott Sheffield and Wendelin Werner for new insights into the topic at the Oberwolfach Workshop Conformal Invariance in Mathematical Physics. Last but not least I am also grateful for Hendrik Adorf's comments on my paper and of course, for Michael Flohr's further advice. 

This research was funded by the Friedrich-Naumann-Stiftung.

\appendix
\section{Appendix: How to Compute ``Fused'' Differential Operators}\label{app:fusion}
In this appendix, we want to show how the additional exponent of the ``fused'' differential operators emerges from general considerations. The whole idea is based on appendix 8.A of \cite{DiFrancesco:1997nk} where the interested reader may find more information on the subject, too.
\subsection*{The Operator Product Expansion}
As already said above, the operator product expansion of two primary fields at $x_0,x_1$ of weights $h_0$ and $h_1$ is given by \begin{equation}
\phi_0(x_0) \phi_1(x_1) \sim
\sum_{h_{(r',s')}} \frac{g_{(r',s')}}{\epsilon^{h_0 +h_1 - h}}
\sum_Y \epsilon^{|Y|} \beta_{Y,{(r',s')}} L_{-Y} \phi_h (x)
\end{equation}
where $\mu = h_0 + h_1 - h_{(r',s')}$ and $g_{(r',s')}$ the coefficient of the three-point function of $\phi_0 (x_0), \phi_1(x_1)$ and $\phi_{h_{(r',s')}}(x)$ (setting $h_{(r',s')}=h$):
\begin{equation}
\frac{g_{(r',s')}}{(x_0-x_1)^{h_0+h_1-h}(x_0-x)^{h_0+h-h_1}(x_1-x)^{h+h_1-h_0} }
= \langle \phi_0 (x_0) \phi_1(x_1) \phi_{h_{(r',s')}} (x) \rangle\,.
\end{equation}
In our case, we have $h_0=h_1=
h_{(2,1)}=\frac{3}{16}\kappa-\frac{1}{2}$, so that $h_{(r',s')}=h_{(1,1)}=0$ or
$h_{(r',s')}=h_{(3,1)}=\frac{\kappa}{2}-1$ since $g_{(r',s')}=0$ for all other values $h_{(r',s')}$. Of course, we can compute the case for 
$h_0=h_1= h_{(1,2)}=\frac{6-\kappa}{2\kappa}$ as well (leading to $h_{(r',s')}=h_{(1,1)}=0$ or $h_{(r',s')}= h_{(1,3)} = \frac{8-\kappa}{\kappa}$), just change $\kappa \rightarrow \frac{16}{\kappa}$ in the computations.

In the following we will concentrate on the $h_{(3,1)}$ branch of fusion since that will be the more interesting case for {\sc SLE} purposes; the identity branch has already been identified with meeting tips of two {\sc SLE} curves that ``annihilate'' and vanish from the scene for all later times -- leaving a setting with two curves less but no perviously unknown features whatsoever \cite{Bauer:2005jt}. Hence, any result we are stating here is always up to the coefficient $g_{(r',s')}$ and the part coming from the identity branch of the fusion process. But since different representations do not mix, the results will nevertheless be exact enough for our purposes.

\subsection*{The Differential Operators}
In the theory of minimal models, the primary fields are degenerate, thus satisfying differential equations. Hence the left hand side of our {\sc OPE} inserted in a correlation function satisfies a so-called null vector differential equation which should be still valid after fusion meaning that we should get differential equations on the {\sc rhs}, too. Hence, for the {\sc lhs}, we have:
\begin{equation}
\langle \phi_0(x_0)
\left[ (\mathcal{L}^n_{-1})^2 (x_1) - t \mathcal{L}^n_{-2}(x_1)
\right] \phi_1(x_1) \phi_3(x_3) \ldots \phi_n (x_n)\rangle = 0
\end{equation}
with
\begin{equation}
\mathcal{L}^n_{-k} (x_i) = \sum_{j\neq i=0}^{n-1} \frac{h(k-1)}{(x_j-x_i)^k} - \frac{1}{(x_j-x_i)^{k-1}}\partial_{x_j} \,.
\end{equation}
and $\mathcal{L}^n_{-1}(x_i) = \partial_{x_i}$. As said above, there should be a corresponding equation for the {\sc rhs} after fusion. Therefore we introduce  $\mathcal{L}_{-r}'(x,\epsilon)$, i.\,e.\ the around
$x = x_1$ expanded differential operators, with $\epsilon=x_0-x_1$ being the deviation of $x_0$ from $x$. Note that we could expand around any other (from $\epsilon$ linearly independent) coordinate since the problem is translational invariant. We just take this special choice here for the computations to be as simple as possible. With some effort we get:
\begin{eqnarray}
\mathcal{L}_{-r}'(x,\epsilon) 
&=& \left.\sum_{j\neq 1}^{n-1} \frac{h(k-1)}{(x_j-x_1)^k} - \frac{1}{(x_j-x_1)^{k-1}}\partial_{x_j} \right|_{x_1=x, x_0-x_1 = \epsilon} \nonumber\\
&=& \frac{h(k-1)}{(x_0-x_1)^k} - \frac{1}{(x_0-x_1)^{k-1}}\partial_{x_0} +
\sum_{j=2}^{n-1} \frac{h(k-1)}{(x_j-x_1)^k} - \frac{1}{(x_j-x_1)^{k-1}}\partial_{x_j} \nonumber\\
&=& \frac{h(k-1)}{\epsilon^k} - \frac{1}{\epsilon^{k-1}}\partial_\epsilon +
\sum_{j=2}^{n-1} \frac{h(k-1)}{(x_j-x)^k} - \frac{1}{(x_j-x)^{k-1}}\partial_{x_j} \nonumber\\
&=& \epsilon^{-r}\left[h_0(r-1)-\epsilon\partial_\epsilon\right] +
\mathcal{L}^{n-1}_{-r}(x)\,,
\end{eqnarray}
and hence
\begin{equation}
\mathcal{L}^n_{-r}(x_1) = \mathcal{L}_{-r}'(x,\epsilon) = \epsilon^{-r} \left[ h_0(r-1) - \epsilon \frac{\text{d}}{\text{d}\epsilon}\right] + \mathcal{L}^{n-1}_{-r} (x)\,.
\end{equation}
Thus, the differential equation on the {\sc rhs} becomes:
\begin{eqnarray}
\left\langle
    \left[
      \left(\mathcal{L}_{-1}' (x,\epsilon)
      \right)^2 - t \mathcal{L}_{-2}'(x,\epsilon)
    \right]
    \left[
    \sum_h \frac{g}{\epsilon^\mu} \sum_Y \epsilon^{Y} \beta_Y
      \left(L_{-Y} \phi_h (x)
      \right)
    \right]
\right. &&\nonumber\\
\left.\times
  \phi_3(x_3) \ldots \phi_{n-1} (x_{n-1})
\right\rangle &=& 0
\end{eqnarray}

\subsection*{The Computation of the Coefficients}
Now we will show how to compute $\beta_{Y,{(r',s')}}$ the from global conformal invariance. It is known that a primary field $\phi_h(z)$ of weight $h$ behaves as follows under conformal transformations $z \rightarrow f(z) = z' = z + \varepsilon(z)$:
\begin{equation}
\phi_h(z) = \left( \frac{\partial f(z)}{\partial z} \right)^h \phi_h(f(z)) \quad \Leftrightarrow \quad \phi_h (z') (\text{d}z')^h = \phi_h(z) (\text{d}z)^h \, .
\end{equation}
Thus, for an infinitesimal transformation, we have:
\begin{equation}
\left(\phi_h(z) + \varepsilon(z) \partial \phi_h(z) \right) (\text{d}z)^h (1+h \varepsilon'(z)) = \phi_h(z)(\text{d}z)^h \, .
\end{equation}
From this follows that for $\varepsilon (z) = \sum_k \varepsilon_k z^{k+1}$ and $\delta_\varepsilon L_{-Y} \phi_h(x) = \varepsilon_k L_k L_{-Y} \phi_h(x)$ after comparing coefficients of powers of $z$, we have for our special situation:
\begin{equation}
\epsilon^{-\mu} \sum_{Y} \epsilon^{|Y|} \beta_Y L_k L_{-Y} \phi_h(x) = \left( h_0 (k+1) \epsilon^k + \epsilon^{k+1} \partial_\epsilon \right) \epsilon^{-\mu} \sum_{Y} \epsilon^{|Y|} \beta_Y L_{-Y} \phi_h(x) \, .
\end{equation}
In other words, all this implies the following rule of thumb:
\begin{itemize}
\item For any level $j$ we take the $j$ equations
\begin{equation}
L_k \phi_h^{(j)}(x) = \left[ h_0 (k+1) + j - k - \mu \right] \phi_h^{(j-k)}(x)
\end{equation}
with $k = 1,\ldots,j$ from the covariance conditions with
\begin{equation}
\phi_h^{(j)} (x) = \sum_{(|Y|)=j} \beta_Y L_{-Y} \phi_h(x) \, .
\end{equation}
\item Moreover, we have to compute the following commutators algebraically:
\begin{equation}
L_k \sum_{|Y|=j} \beta_Y L_{-Y} \phi_h(x) = \sum_{|Y|=j} \beta_Y \left[L_k , L_{-Y}\right] \phi_h (x) ,
\end{equation}
(since $\phi_h (x)$ is a primary field, the action of positive generators of the {\sc Virasoro}-algebra vanishes).
\end{itemize}
Afterwards we compare the respective coefficents of $\phi_h^{(j-k)}(x)$ and hence get the $\beta_Y$ that depend on the choice of coordinate change $(x_0,x_1) \rightarrow (x,\epsilon)$.
\subsection*{The New Differential Equation}
Hence the new differential operators are:
\begin{eqnarray}
\mathcal{L}_{-1}'
&=& \epsilon^{-1} \left[ h_0(1-1) - \epsilon \frac{\text{d}}{\text{d}\epsilon}\right] + \mathcal{L}_{-1} (x) \nonumber\\
&=& - \frac{\text{d}}{\text{d}\epsilon} + \mathcal{L}_{-1}(x) \\
\mathcal{L}_{-2}' &=& \epsilon^{-2} \left[ h_0(2-1) - \epsilon \frac{\text{d}}{\text{d}\epsilon}\right] + \mathcal{L}_{-2} (x) \nonumber\\
&=& \frac{h_0}{\epsilon^2} - \frac{1}{\epsilon} \frac{\text{d}}{\text{d}\epsilon} + \mathcal{L}_{-2} (x)
\end{eqnarray}
with which we can state the differential equation for the fused correlation function:
\begin{eqnarray}
&&\left\langle \left[\left(\frac{\text{d}^2}{\text{d}\epsilon^2} - 2 \frac{\text{d}}{\text{d}\epsilon} \mathcal{L}_{-1} (x) + \mathcal{L}_{-1}^2(x) \right) \right.\right.\nonumber\\
&&- \left.\left.t \left( \frac{h_0}{\epsilon^2} - \frac{1}{\epsilon} \frac{\text{d}}{\text{d}\epsilon} + \mathcal{L}_{-2} (x) \right)\right] \epsilon^{-\mu} \sum_j \epsilon^j \phi_h^{(j)}\times\phi_2(x_2) \ldots \phi_h(x_n) \right\rangle = 0 \nonumber
\end{eqnarray}
In the following, we set $h_{(3,1)} = h$. After executing the derivative with respect to $\epsilon$ and ordering the coefficients of the various powers of $\epsilon$, we get:
\begin{eqnarray}
0 &=& \left[ (j+2-\mu)(j+2-\mu-1) - t (h_0 - (j+2-\mu) \right]
\phi_h^{(j+2)} \nonumber\\
&\quad& - 2(j+1-\mu) L_{-1} \phi_h^{(j+1)} + \left( L_{-1}^2 -
t L_{-2} \right) \phi_h^{(j)}
\end{eqnarray}
\subsection*{The Reason for $\epsilon^{1-\mu}$}
Now we want to get a level three null vector condition. For dimensional reasons, it is quite obvious that we have to take the $j=1$ term which gives us an additional factor of $\epsilon$: after finishing the computations, the {\sc rhs} will be proportional to $\epsilon^{j-\mu}$ times the differential operator on level three times a correlation function involving $\phi_h$ which has the same transformational properties under conformal transformaions as the level two differential operator acting on a correlation function involving $\phi_{h_0}$ and $\phi_{h_1}$. This is the crucial point why we have an additional distance factor in our equations in the main part of this paper.

The interested reader may question why it is sufficient to just compute the $j=1$ part of the {\sc OPE}. The shortest argument possible can be stated as follows: since we already know the outcome of fusion -- for the fields as well as for the differential operator, we know, too, that the other terms have to vanish as a we know that the differential equations have to be fullfilled before and after fusion. A more elaborate reasoning may be found in section appendix 8.A.3 of \cite{DiFrancesco:1997nk}.

\subsection*{Computing the Coefficients}
Setting $j=1$, we get:
\begin{eqnarray}
0&=& \underbrace{\left[ (3-2 h_0+h)(2-2 h_0+h) - t (3h_0 - 3 -h)
\right]}_{V^1_1} \phi_h^{(3)} \\
&\quad&
\underbrace{- 2(2-2h_0+h)}_{V^1_2} L_{-1} \phi_h^{(2)} + L_{-1}^2\phi_h^{(1)} 
\underbrace{- t}_{V^1_3} L_{-2}
\phi_h^{(1)}\nonumber\\
&=& \left[V_1^1 \left( \beta^3_{111} L_{-1}^3 + \beta^3_{12} L_{-1} L_{-2} +  \beta^3_{21} L_{-2} L_{-1} + \beta^3_3 L_{-3}\right) \right. \\
&\quad&\left.+ V^1_2 L_{-1} \left(\beta^3_{11} L_{-1^2} + \beta^3_2 L_{-2} \right) + L_{-1}^2 \beta^3_1 L_{-1} + V^1_3 L_{-2} \beta^3_1 L_{-1}\right] \phi_h(x) \nonumber\\
&=& K^3_1 L_{-1}^3 + K^3_{12} L_{-1}L_{-2} + K^3_{21} L_{-2} L_{-1} + K^3_3 L_{-3}
\end{eqnarray} with
\begin{eqnarray}
K^1_1&:\,=& V^3_1 \beta^3_{111} + V^3_2 \beta^3_{11} + \beta^3_1 \\
K^1_{12}&:\,=& V^3_1 \beta^3_{12} + V^3_2 \beta^1_2\\
K^1_{21}&:\,=& V^3_1 \beta^3_{21} + V^3_3 \beta^1_1\\
K^1_3&:\,=& V^3_1 \beta^3_3
\end{eqnarray}
using the commutator $L_{-1} L_{-2} = L_{-2} L_{-1} + L_{-3} $.

Knowing the form of the algebraic level three null operator:
\begin{equation}
L_{-1}^3 - (h+1)( L_{-2}L_{-1}+ L_{-1}L_{-2}) + (h+1)^2L_{-3}\, ,
\end{equation}
we only have to compare the coefficients:
\begin{eqnarray}
0&=&K_1^3 L_{-1}^3 + K_{12}^3 L_{-1}L_{-2} + K_{21}^3 L_{-2}L_{-1} +
K_3^3 L_{-3} \nonumber\\
&\propto& L_{-1}^3-(h+1) L_{-2}L_{-1} -(h+1)L_{-1}L_{-2} +
(h+2)(h+1) L_{-3}
\end{eqnarray}
wherein the proportionality factor $K^3_1$ can in principle dpend on $h$. Hence all that is left to be shown is:
\begin{eqnarray}
-(h+1) &=& \frac{K_{12}^3}{K_1^3} \\
-(h+1) &=& \frac{K_{21}^3}{K_1^3} \\
(h+1)^2 &=& \frac{K_3^3}{K_1^3}
\end{eqnarray}
for which we have to compute the exact values of the 
$\beta_Y$. Therefore we will use the following reexpressions:
\begin{eqnarray}
t &=& \frac{h+1}{2}\\
c &=& -\frac{(3h-1)(h-2)}{h+1}\\
h_0 &=& h_1 = \frac{1}{8} (3h-1)\,.
\end{eqnarray}

Now we will use our rule of thumb stated above:
\begin{itemize}
\item[$j=1$]
from covariance:
\begin{eqnarray}
L_1 \phi_h^{(1)}(x) &=& \left[ h_0 (1+1) + 1 - 1 - (2h_0-h) \right] \phi_h^{(1-1)}(x) \nonumber\\
&=& h \phi_h (x)
\end{eqnarray}
algebraically:
\begin{eqnarray}
L_1 \phi_h^{(1)} (x) &=& L_1 L_{-1} \beta^3_1 \phi_h (x) \nonumber\\
&=& 2 h \beta^3_1 \phi_h (x)
\end{eqnarray}
Hence it follows $\beta^3_1 = 1/2$.
\item[$j=2$]

from covariance:
\begin{eqnarray}
L_1 \phi_h^{(2)}(x) &=& \left[ h_0 (1+1) + 2 - 1 - (2 h_0 - h) \right] \phi_h^{(2-1)}(x) \nonumber\\
&=& (h+1) \phi_h^{(1)} (x) \nonumber\\
&=& \frac{h+1}{2} L_{-1} \phi_h (x) \\
L_2 \phi_h^{(2)}(x) &=& \left[ h_0 (2+1) + 2 - 2 - (2 h_0 - h) \right] \phi_h^{(2-2)}(x) \nonumber\\
&=& (h+h_0) \phi_h (x)
\end{eqnarray}
algebraically:
\begin{eqnarray}
L_1 \phi_h^{(2)}(x) &=& L_1 \left(\beta^3_{11} L_{-1}^2 + \beta^3_2 L_{-2} \right) \phi_h (x) \nonumber\\
&=& \left( 2(2h+1) \beta^3_{11} + 3 \beta^3_2 \right) L_{-1} \phi_h(x) \\
L_2 \phi_h^{(2)}(x) &=& L_2 \left(\beta^3_{11} L_{-1}^2 + \beta^3_2 L_{-2} \right) \phi_h (x) \nonumber\\
&=& \left( 6h \beta^3_{11} + (4h + c/2) \beta^3_2 \right) \phi_h(x)
\end{eqnarray}
comparing the coefficients, it follows:
\begin{eqnarray}
\beta_{11}^3 &=& \frac{1}{8}\frac{h+1}{h+2}\\
\beta_2^3 &=& \frac{1}{4}\frac{h+1}{h+2}
\end{eqnarray}
\item[$j=3$]

from covariance:
\begin{eqnarray}
L_1 \phi_h^{(3)}(x) &=& \left[ h_0 (1+1) + 3 - 1 - (2 h_0 - h) \right] \phi_h^{(3-1)}(x) \nonumber\\
&=& (h+2) \phi_h^{(2)} (x) \nonumber\\
&=& (h+2) \left(\beta^3_{11} L_{-1}^2 + \beta^3_2 L_{-2} \right) \phi_h (x)  (x) \\
L_2 \phi_h^{(3)}(x) &=& \left[ h_0 (2+1) + 3 - 2 - (2 h_0 - h) \right] \phi_h^{(3-2)}(x) \nonumber\\
&=& (h+1+h_0) \phi_h^{(1)} \nonumber\\
&=& \frac{h+1+h_0}{2} L_{-1} \phi_h (x) \\
L_3 \phi_h^{(3)}(x) &=& \left[ h_0 (3+1) + 3 - 3 - (2 h_0 - h) \right] \phi_h^{(3-3)}(x) \nonumber\\
&=& (h+2h_0) \phi_h (x)
\end{eqnarray}
algebraically:
\begin{eqnarray}
L_1 \phi_h^{(3)}(x) &=& L_1 \left(L_{-1}^3 \beta^3_{111}+L_{-1}L_{-2} \beta^3_{12} + L_{-2} L_{-1} \beta^3_{21} + L_{-3} \beta^3_3 \right) \phi_h(x) \nonumber\\
&=&
\left(\left[ 6(h+1) L_{-1}^2\right]  \beta^3_{111}+ \left[ 3L_{-1}^2 + 2(h+2) L_{-2} \right]\beta^3_{12} \right.\nonumber\\
&\quad& \left.+ \left[ 2hL_{-2} + 3L_{-1}^2 \right]  \beta^3_{21} + \left[ 4 L_{-2}\right]  \beta^3_3 \right) \phi_h(x) \nonumber\\
&=& \left( \left[ 6(h+1) \beta^3_{111} + 3 \beta^3_{12} + 3 \beta^3_{21} \right] L_{-1}^2 \right.\nonumber\\
&\quad& \left.+ \left[ 2(h+2) \beta^3_{12} + 2h \beta^3_{21} + 4 \beta^3_3 \right] L_{-2} \right) \phi_h (x) \\
L_2 \phi_h^{(3)}(x) &=& L_2 \left(L_{-1}^3 \beta^3_{111}+L_{-1}L_{-2} \beta^3_{12}  + L_{-2} L_{-1} \beta^3_{21} + L_{-3} \beta^3_3 \right) \phi_h(x) \nonumber\\
&=&
\left(\left[6(3h+1) \right] \beta^3_{111}+ \left[ 5+4(h+1)+\frac{c}{2}\right]\beta^3_{12} \right.\nonumber\\
&\quad& \left.+  \left[ 4(h+1) +c/2 \right] \beta^3_{21} + \left[ 5 \right] \beta_3 \right) L_{-1} \phi_h(x) \\
L_3 \phi_h^{(3)}(x) &=& L_3 \left(L_{-1}^3 \beta^3_{111} + L_{-1}L_{-2} \beta^3_{12}  + L_{-2} L_{-1} \beta^3_{21} + L_{-3} \beta^3_3 \right) \phi_h(x) \nonumber\\
&=& \left( 24h \beta^3_{111} + (16h+2c)\beta^3_{12} + 10h \beta^3_{21} +(6h+2c) \beta^3_3 \right) \phi_h(x) \nonumber\\
&\quad&
\end{eqnarray}
comparing the coefficients, it follows:
\begin{eqnarray}
\beta^3_{111} &=& \text{free}\\
\beta^3_{12}&=&h(h+1) \beta^3_{111} - \beta^3_3 - \frac{(h-3)(h-1)}{48}\\
\beta^3_{21}&=& -(h+2)(h+1) \beta^3_{111} + \beta^3_3 + \frac{(h-3)(h-1)}{48}\\
\beta^3_3 &=&\text{fre}
\end{eqnarray}
\end{itemize}
Introducing
\begin{eqnarray}
I^3_2 := \frac{K^3_{12} + K^3_{21}}{K^3_1} =!= -2(h+1)\\
I^3_3 := \frac{K^3_{12} + K^3_3}{K^3_1} =!= h(h+1)
\end{eqnarray}
we get
\begin{eqnarray}
K_1^3 &=& \frac{1}{16(h+2)}\left[ 48 \beta^3_{111} (h+2)(h+3) - (h^2+2h-7) \right]\\
K_{12}^3 
&=& h(h+1)K^3_1 - 3 \beta^3_3 (h+3)\\
K_{21}^3 
&=& -(h+2)(h+1) K^3_1 + 3 (h+3) \beta^3_3\\
K_3^3 &=& 3(h+3) \beta^3_3
\end{eqnarray}
Thus the prefactors become:
\begin{eqnarray}
H^3_{12} 
&=& h(h+1) - \frac{48 \beta^3_3 (h+2)(h+3)}{48 \beta^3_{111} (h+2)(h+3) - (h^2+2h-7)}\\
H^3_{21} 
&=& -(h+2)(h+1) + \frac{48\beta^3_3 (h+2)(h+3)}{48 \beta^3_{111} (h+2)(h+3) - (h^2+2h-7)}\\
H^3_3 &=& \frac{48 \beta^3_3 (h+2)(h+3)}{48 \beta^3_{111} (h+2)(h+3) - (h^2+2h-7)}
\end{eqnarray}
which leads us obviously to
\begin{eqnarray}
I^3_2 = H^3_{12} + H^3_{12}&=& -2(h+1) \\
I^3_3 = H^3_{12} + H^3_{3}&=& h(h+1)
\end{eqnarray}
hence the level two null vector condition on a primary  $\phi_{(2,1)}$ field translates to a level three null vector equation on a primary $\phi_{(3,1)}$ field after fusion with another $\phi_{(2,1)}$ field.
\small
\bibliography{Bibliography}
\end{document}